\documentclass[trackchanges]{aastex631}
\usepackage{amsmath}
\usepackage{bm}
\usepackage[shortlabels]{enumitem}

\shorttitle{Pulsar Average Emission properties from Solitons}

\begin{document}

\title{Pulsar Coherent Radio Emission from Solitons : Average Emission Properties}


\author[0000-0003-1824-4487]{Rahul Basu}
\affiliation{Janusz Gil Institute of Astronomy, University of Zielona G\'ora, ul. Szafrana 2, 65-516 Zielona G\'ora, Poland.}

\author[0000-0002-9142-9835]{Dipanjan Mitra}
\affiliation{National Centre for Radio Astrophysics, Tata Institute of Fundamental Research, Pune 411007, India.}
\affiliation{Janusz Gil Institute of Astronomy, University of Zielona G\'ora, ul. Szafrana 2, 65-516 Zielona G\'ora, Poland.}

\author[0000-0003-1879-1659]{George I. Melikidze}
\affiliation{Janusz Gil Institute of Astronomy, University of Zielona G\'ora, ul. Szafrana 2, 65-516 Zielona G\'ora, Poland.}
\affiliation{Evgeni Kharadze Georgian National Astrophysical Observatory, 0301 Abastumani, Georgia.}

\author[0000-0003-0338-7506]{Krzysztof Maciesiak}
\affiliation{Janusz Gil Institute of Astronomy, University of Zielona G\'ora, ul. Szafrana 2, 65-516 Zielona G\'ora, Poland.}

\begin{abstract}
Observations have established that coherent radio emission from pulsars arise 
at few hundred kilometers above stellar surface. Recent polarization studies 
have further demonstrated that plasma instabilities are necessary for charge 
bunching that gives rise to coherent emission. The formation of charged 
solitons in the electron-positron plasma is the only known bunching mechanism 
that can be realised at these heights. More than five decades of observations 
have revealed a number of emission features that should emerge from any valid 
radio emission mechanism. We have carried out numerical calculations to find 
the features of average emission from curvature radiation due to charged 
solitons. The characteristic curvature radiation spectrum has been updated from
the well known one-dimensional dependence into a general two-dimensional form, 
and contribution from each soliton along observer's line of sight (LOS) has 
been added to reproduce the pulsar emission. The outflowing plasma is formed by
sparking discharges above the stellar surface that are located within 
concentric rings resembling the core-cone emission beam, and uniform 
distribution of solitons along any LOS has been assumed. The observed effects 
of radius to frequency mapping, where the lower frequency emission originates 
from higher altitudes, is seen in this setup. The power law spectrum and 
relative steepening of the core spectra with respect to the cones also emerges.
The estimated polarization position angle reflects the geometrical 
configuration of pulsars as expected. These studies demonstrate the efficacy of
coherent curvature radiation from charged solitons to reproduce the average 
observational features of pulsars. 
\end{abstract}


\section{Introduction}
The origin of coherent radio emission from pulsars has remained an unresolved 
topic despite more than five decades of intensive research \citep[see, 
e.g.][for a review]{M17,MBM24a}. The highly magnetized ($B\sim10^{12}$ G), 
fast-rotating neutron star (rotation period $P\sim1$ second) with a 
superconducting interior causes redistribution of surface charge density and 
generates a strong induced electric field ($E \sim 10^{12}$ V/m). In these 
environments, the creation of copious electron-positron pairs naturally fills 
the pulsar magnetosphere with dense plasma, which co-rotates with the star. The
pulsar magnetosphere can be divided into closed and open magnetic field line 
regions. In the closed field line region, a steady state charge distribution 
with a density, known as the Goldreich-Julian density, $\rho_{GJ}$ = 
$\Omega\cdot B/2\pi c$, where $\Omega = 2\pi/P$, screens the potential 
difference \citep{GJ69}. The open field line region encompasses an area above 
the neutron star surface, centered around the magnetic axis, known as the polar
cap. The co-rotation speed exceeds the speed of light in the open field line 
region beyond a distance known as the light cylinder radius, $R_{LC}=P\cdot 
c/2\pi$, and hence the steady state configuration cannot be reached. Instead, a
non-stationary outflow of plasma is expected along the open field lines 
sustained by electron-positron pair production in presence of large magnetic 
field from a tiny inner acceleration region (IAR) above the polar cap 
\citep{S71,RS75}. The observed coherent radio emission is expected to emerge 
from non-linear instabilities in the outflowing plasma resulting in charge 
bunching \citep{MP80,MGP00,LMM18,RMM22b}. The principal challenges in 
understanding the origin of radio emission include finding a consistent model 
for the generation of plasma in the pulsar magnetosphere, and a unique coherent
emission mechanism due to instabilities in plasma.

In recent years two significant developments have made it possible to 
understand the physical processes leading to the radio emission in pulsars. The
first involves identifying the plasma generation from a partially screened gap 
(PSG) above the polar cap \citep{GMG03}, and the second being the demonstration
that stable charge bunches in the form of solitons can develop due to 
two-stream instability in the pulsar plasma \citep{MP80,AM98,MGP00} and give 
rise to coherent curvature radiation (CCR). The radio emission from pulsars is 
seen in shorts burst that usually are limited to less than 10\% of the rotation
period. Each single pulse comprises of or more distinct structures also known 
as subpulses. The average profile formed after adding several thousand single 
pulses is unique to each pulsar and shows the presence of more than one 
component. The presence of isolated sparking discharges in the IAR, due
to cascading electron-positron pair production, was postulated by \cite{RS75} 
to explain the origin of the observed components and subpulses, and can also 
explain the phenomenon of subpulse drifting seen in the single pulse sequence
of several pulsars \citep{WES06,BMM16}. However, subsequent studies have shown 
that in a purely vacuum gap above the polar cap, the isolated sparks with 
discharge free zones in between are unstable \citep{CR80,CPT24}. The problem of
forming stable sparks has been successfully addressed by the PSG model that 
considers a non-dipolar polar cap whose surface is heated to temperatures 
$\sim10^6$ K by back-flowing particles during sparking discharges. This results 
in a steady outflow of positively charged ions from the surface that screens 
the potential difference in the gap region. Sparks are formed when the surface 
cools below the critical temperature for ionic outflow and is a mechanism of 
thermostatic regulation of the actual surface temperature \citep[see][]{GMG03}.
The sparks in the PSG model are formed in tightly packed configurations, 
without any discharge free zones in between, in concentric rings with the 
outermost ring closely bordering the polar cap boundary along with a central 
spark \citep{BMM20b,BMM22b}. The presence of non-dipolar fields and PSG nature 
of the IAR has been found to be consistent with detailed observations of 
subpulse drifting \citep{BMM19,BMM23a}. These also form the basic requirements 
for sustaining the radio emission in long period single component pulsars like 
J2144$-$3933, that are at the boundary of the death line in the pulsar 
population \citep{MBM20}. The PSG nature of IAR is also necessary to explain 
the origin of thermal X-ray emission from the stellar surface \citep{SGZ17,
AM19,SG20,PM20}.
 
The primary plasma is produced during sparking and after they leave the IAR, 
the pair production continues outside the gap, resulting in dense clouds of 
secondary electron-positron pair plasma. Additional details about the radio 
emission mechanism emerges from the polarization measurements. Pulsar emission 
has high levels of linear polarization where the polarization position angle 
(PPA) has a S-shaped variation across the pulsar profile. This behaviour has 
been interpreted using the rotating vector model \citep[RVM,][]{RC69}, where 
the position angle of the projected electric field vector along the LOS changes
due to the rotation of the pulsar. The RVM naturally favours the curvature 
radiation from ultra-relativistic charge bunches, moving along curved magnetic 
field lines as a mechanism for the radio emission. The PPA deviates from the 
RVM nature as a result of shifts introduced by relativistic 
aberration-retardation effects, arising due to co-rotation of the plasma 
\citep{BCW91,D08}. The delay-radius shifts is measured in the average profile 
between the profile center and the steepest gradient (SG) point of the PPA, and
is proportional to the radio emission height. This method has been used to 
estimate the emission height in a large sample of pulsars, and have constrained
radio emission to originate below 10\% of the light cylinder radius, at a 
distance of 100 -- 1000 km from the surface \citep{M17,MMB23b}. 

Another important constrain on the radio emission mechanism comes from the 
orthogonal polarization modes (OPM) of the PPA seen in several pulsars, with 
two parallel RVM like tracks in the single pulse PPA distributions separated by
90$\degr$ phase difference \citep{BR80,GL95}. The presence of OPM in the PPA 
makes the vacuum curvature radiation models untenable and suggests the 
propagation of radio emission through the ambient plasma affects the radio 
emission properties. In the strongly magnetized, homogeneous, pair plasma of 
pulsars the OPMs correspond to the orthogonally polarized eigenmodes, the 
ordinary (O-mode) and extraordinary (X-mode) waves. The waves are excited by a 
suitable emission mechanism, propagate in the ambient plasma and eventually 
detach from the medium to reach the observer \citep[see][for a
review]{MBM24a}. Linear electrostatic Langmuir waves are formed in the 
non-stationary plasma due to two stream instability at the radio emission 
heights \citep{U87,AM98,RMM20}. The modulational instability developing in the 
Langmuir waves form envelopes of charge separated solitons that are stable 
enough for sufficient durations to emit CCR \citep{MP80,MGP00,GLM04,LMM18,
RMM22b}.

The CCR from charged solitons will be a relevant emission mechanism in pulsars 
only if it can explain the large number of observational features \citep[see 
e.g.][for a discussion]{M17,MBM24a}. In this paper our aim is to carry out 
detailed numerical estimates of the different emission features from solitons 
forming in the pulsar plasma. A two dimensional distribution of solitons in 
the open field line region of pulsars has been used to estimate the average 
spectra and the relative variations across the emission window by 
\cite{BMM22a}. We have updated the soliton curvature radiation intensity 
spectrum from an one dimensional angular dependence to a more general two 
dimensional form as shown in section \ref{sec:solspec}. We have also used a 
three dimensional distribution of solitons in the open field line region and 
estimated the average emission properties by adding the contributions of all 
solitons along every line of sight. These estimates are used to investigate the
average radio emission features in section \ref{sec:Avgem} : the average 
profile and its evolution with frequency as well as emission geometry, the 
location of the emission within the open field line region, the PPA swing 
across the profile and the spectral nature of pulsar emission due to CCR from 
charged solitons. Finally, in section \ref{sec:dis} we discuss the 
effectiveness of curvature radiation from charged solitons to understand the 
observed average radio emission properties from pulsars, and the role of 
dynamic plasma effects on the time varying emission features, that will be 
explored in future works.

\section{Curvature radiation from Solitons in Pulsar Magnetosphere} \label{sec:solspec}

\subsection{Two Dimensional Intensity Pattern of Soliton}
The outflowing non-stationary plasma in the pulsar emission region comprises of
secondary electron-positron pair plasma clouds with typical Lorentz factors, 
$\gamma_s\sim100$, and high density, $n_s\sim\kappa n_{\rm GJ}$, with 
$\kappa\geq10^4$ \citep{S71,TH19}. In this plasma the two stream instability, 
induced either by overlapping of plasma particles from successive clouds 
\citep{U87,AM98} or by the longitudinal drift \citep{RMM20}, trigger the strong
Langmuir turbulence. The non-linear evolution of high-amplitude Langmuir wave 
packets results in the formation of stable non-linear waves, i.e., solitons. 
The charge separation in the magnetosphere introduces a difference in the 
energy distribution of electrons and positrons such that, and the relativistic 
mass difference between the two species produces inertial charge separation 
within the soliton envelope. The formation and stability of solitons in the 
pulsar plasma has been demonstrated in a number of works \citep{MGP00,LMM18,
RMM22b}. In this paper we make use of the soliton characteristics introduced in
\cite{MGP00} to estimate the emission properties of pulsars.

A quantitative estimate of the three dimensional structure of solitons in the 
pulsar plasma remains unavailable, but we can make qualitative estimates based 
on requirements for the coherent radio emission. The coherence condition 
requires the dimensions of solitons to be smaller than half-wavelength 
($\lambda/2$) of emitted radiation. Solitons with isotropic shapes in the 
plasma frame, having typical size $d$, move with relativistic speeds along the
magnetic field lines where group velocity of the soliton wave packet is 
specified by the Lorentz factor $\gamma_o=y\gamma_s$, with $y\sim2-3$. In the
pulsar frame of reference the length contraction along the magnetic field lines
imply that the typical soliton size is $d'\sim d/\gamma_o \lesssim \lambda/2$. 
The transverse size of solitons remain $d \lesssim \gamma_o\lambda/2$, and in 
the pulsar frame the solitons resemble pancake like shapes \citep{GLM04}. Along
the direction of motion a soliton wave packet has been modelled in its simplest
form as a system of three charges $Q_s/2$, $-Q_s$ and $Q_s/2$ separated by 
a characteristic length $\Delta_s$ \citep[see Fig 3 in][]{MGP00}, that can be 
estimated as :
\begin{equation}
\Delta_s \sim \Delta_o \gamma_2^{0.5} \kappa_4^{-0.5} r_{50}^{1.5} P^{0.25} \dot{P}_{-15}^{-0.25}.
\label{eq:sol_len}
\end{equation}
Here $\Delta_o\sim0.1-1$ m, is the characteristic soliton length, $\gamma_2 = 
\gamma_s/100$, $\kappa_4=\kappa/10^4$, $r_{50}=r/50R_s$, $r$ being the emission
height, $R_s=10$ km, the neutron star radius, and $\dot{P}_{-15}$ is the period
derivative of pulsar in units of $10^{-15}$ s/s. The value of the central 
charge density in solitons is $\rho_s\sim10\rho_{GJ}$, and hence the total 
charge can be estimated as $Q_s=\rho_s V_s$, where $V_s = S_\perp \Delta_s$, 
and $S_\perp \propto r^2$.

In the case of a charge separated soliton, as considered above, the radiation 
intensity, $\mathcal{I}_s$, at a frequency $\omega$=2$\pi\nu$, can be 
determined using the Stokes parameters corresponding to single particle 
curvature radiation spectra, $\mathcal{F} = (I, Q, U, V)$, and has the form 
\citep{MGP00} :
\begin{equation}
\mathcal{I}_s(\omega, \theta, \phi, \gamma_o) = Q_s^2~\mathcal{F}(\omega, \theta, \phi, \gamma_o) \left[1 - \cos{\left(a_s\frac{\omega}{\omega_o}\right)}\right]^2.
\label{eq:sol_spec}
\end{equation}
Here, $\theta$ and $\phi$ are the angular separations measured with respect to 
the tangential direction along the magnetic field line at the soliton center 
(see Fig.~\ref{fig:curv_cord} in appendix \ref{app:curv_spec}), $\omega_o = 
\frac{3}{2} \gamma_o^3 c/\mathcal{R}$, is the characteristic frequency of 
curvature radiation with $\mathcal{R}$ being the radius of curvature of the 
magnetic field line (see appendix \ref{app:radcurv}), and $a_s = 
(\Delta_s/\mathcal{R})\gamma_o^3$. 

\begin{figure}
\gridline{\fig{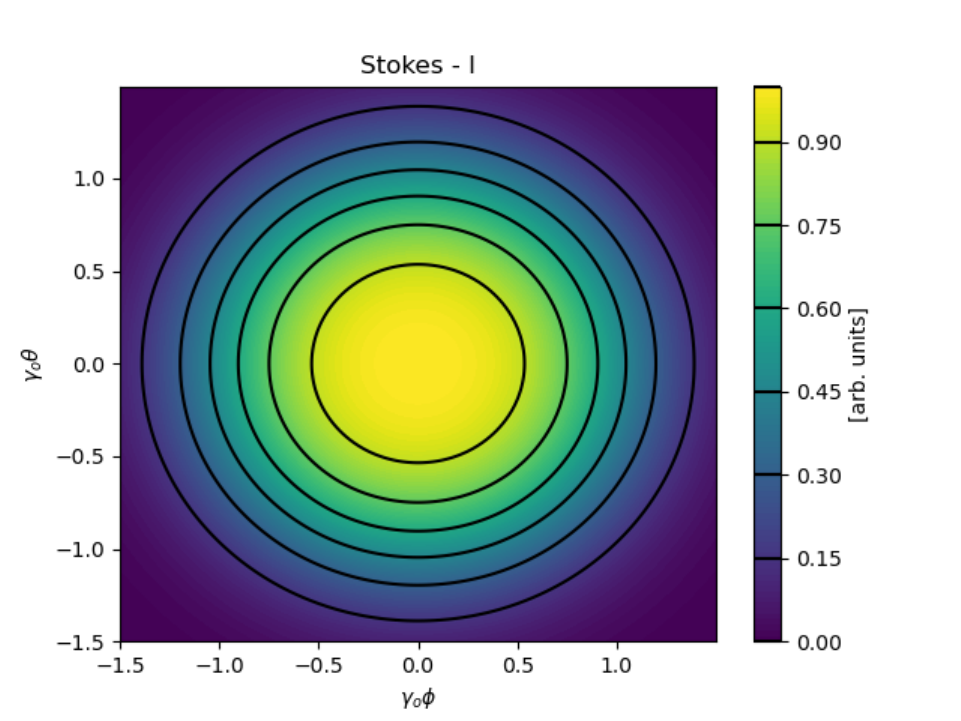}{0.45\textwidth}{(a)}
          \fig{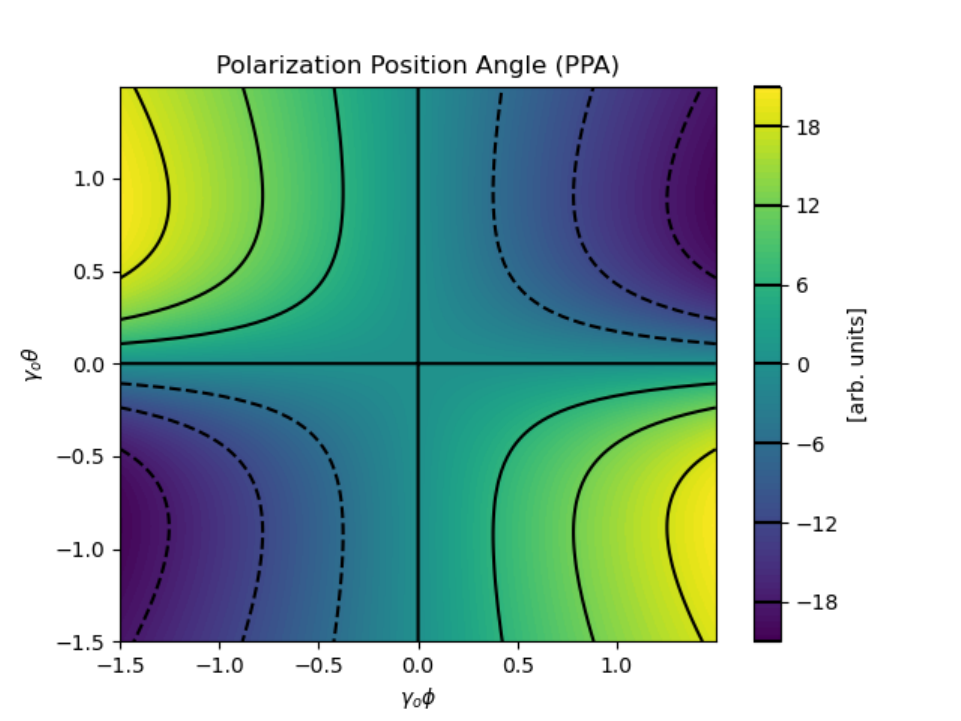}{0.45\textwidth}{(b)}
         }
\gridline{\fig{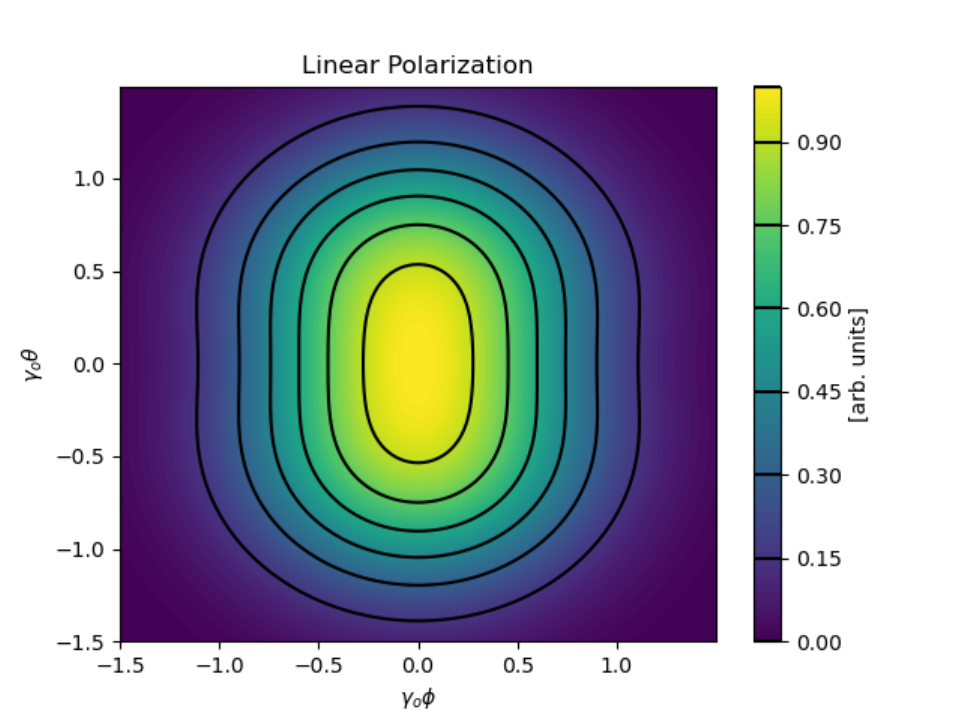}{0.45\textwidth}{(c)}
	  \fig{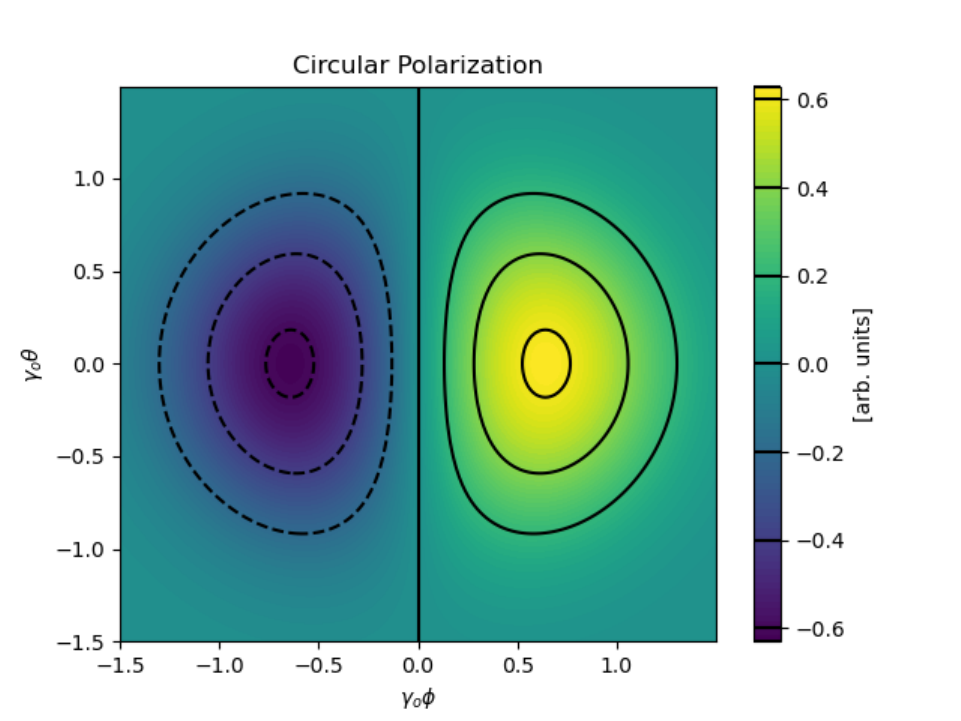}{0.45\textwidth}{(d)}
         }
\gridline{\fig{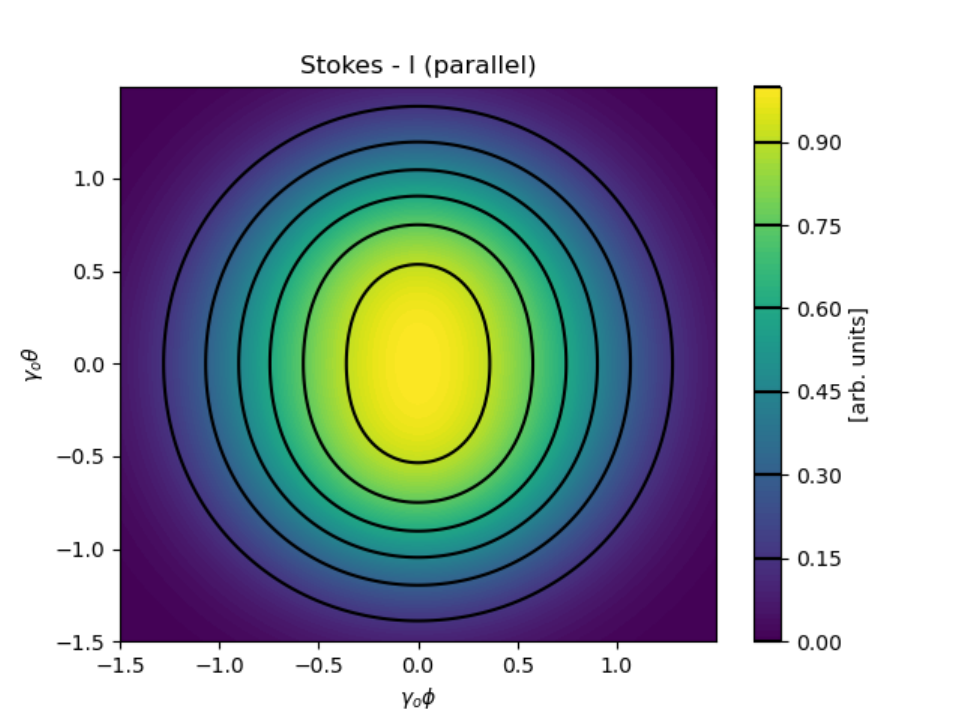}{0.45\textwidth}{(e)}
          \fig{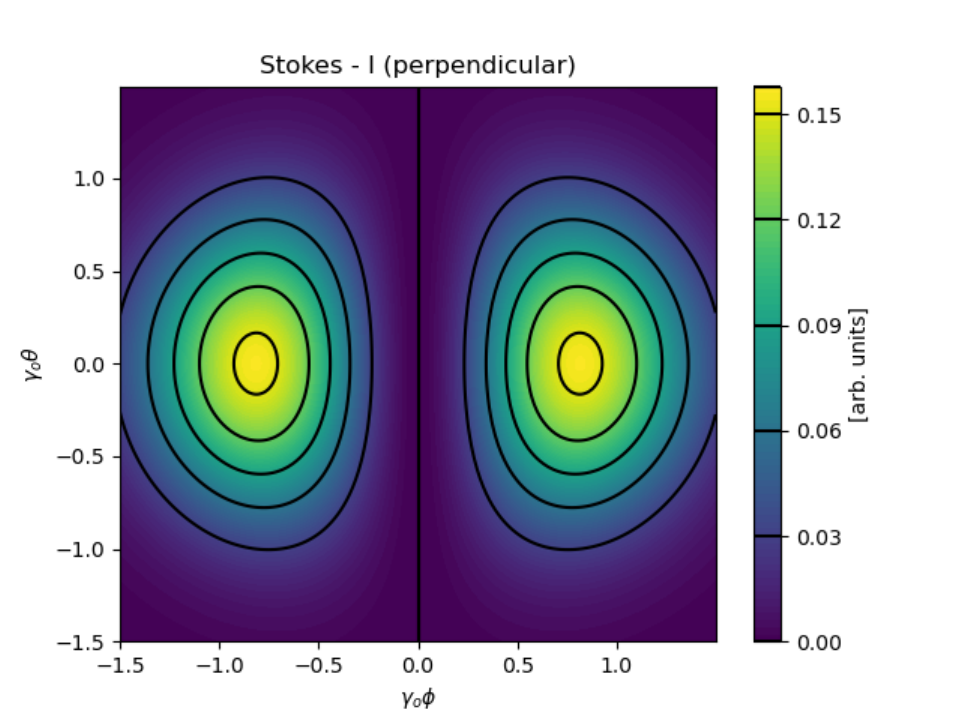}{0.45\textwidth}{(f)}
         }
\caption{The two dimensional distribution of the curvature radiation from a
charged soliton estimated at frequency $\omega = \omega_c$, as a function of 
normalized angular co-ordinates, $\gamma_o\theta$ and $\gamma_o\phi$, measured 
with respect to the tangent vector of the magnetic field line at the center of 
the soliton. The estimates for vacuum propagation of electromagnetic waves are 
shown : (a) total intensity (I) (b) polarization position angle (PPA) (c) 
linear polarization (L) and (d) circular polarization (V). In pulsar plasma the 
spectra gets separated into two parts : (e) parallel mode and (f) perpendicular
mode.
\label{fig:sol_prop}}
\end{figure}

The curvature radiation from a relativistic charged particle moving along a
curved field line is a text book problem \citep[see for e.g.][]{J98}, where
the spectral variation is usually estimated over a single angular co-ordinate.
Although the emission is highly beamed and restricted within a narrow emission 
cone around the propagation direction, we still require an estimate along both 
angular co-ordinates to suitably add up the contributions of individual 
solitons to the total emission. A derivation of the two-dimensional forms of 
the curvature radiation electric fields as a function of $\theta$ and $\phi$, 
under small angle approximation, is shown in appendix \ref{app:curv_spec}. If 
we consider propagation in vacuum, i.e. the electromagnetic modes generated by 
the solitons detach immediately after they are formed, the two orthogonal 
polarization modes remain coupled and the four Stokes parameters have the form :
\begin{eqnarray}
I\left(\frac{\omega}{\omega_o},\theta,\phi, \gamma_o\right) & = & \gamma_o^2 \left(\frac{\omega}{\omega_o}\right)^2~(1 + \gamma_o^2\theta^2 + \gamma_o^2\phi^2)^2~\left[K_{2/3}^2(\xi) + \left(\frac{\gamma_o^2\theta^2 + \gamma_o^2\phi^2}{1 + \gamma_o^2\theta^2 + \gamma_o^2\phi^2}\right)K_{1/3}^2(\xi)\right], \label{eq:curv_spec1}\\
Q\left(\frac{\omega}{\omega_o},\theta,\phi, \gamma_o\right) & = & \gamma_o^2 \left(\frac{\omega}{\omega_o}\right)^2~(1 + \gamma_o^2\theta^2 + \gamma_o^2\phi^2)^2~\left[K_{2/3}^2(\xi) + \left(\frac{\gamma_o^2\theta^2 - \gamma_o^2\phi^2}{1 + \gamma_o^2\theta^2 + \gamma_o^2\phi^2}\right)K_{1/3}^2(\xi)\right], \label{eq:curv_spec2} \\
U\left(\frac{\omega}{\omega_o},\theta,\phi, \gamma_o\right) & = & -2 \gamma_o^2 \left(\frac{\omega}{\omega_o}\right)^2~(1 + \gamma_o^2\theta^2 + \gamma_o^2\phi^2) (\gamma_o^2\theta\phi) K_{1/3}^2(\xi), \label{eq:curv_spec3} \\
V\left(\frac{\omega}{\omega_o},\theta,\phi, \gamma_o\right) & = & 2 \gamma_o^2 \left(\frac{\omega}{\omega_o}\right)^2~(1 + \gamma_o^2\theta^2 + \gamma_o^2\phi^2)^{3/2}(\gamma_o\phi) K_{1/3} K_{2/3}(\xi), \label{eq:curv_spec4}
\end{eqnarray}
here $\xi = \frac{\omega}{2\omega_o} \left(1 + \gamma_o^2\theta^2 +
\gamma_o^2\phi^2\right)^{3/2}$ and $K_{1/3}(\xi)$ and $K_{2/3}(\xi)$ are the 
modified Bessel functions. The linear polarization is obtained as $L = 
\sqrt{Q^2+U^2}$, while the polarization position angle, PPA = 
$0.5\tan^{-1}(U/Q)$. The equivalent quantities for the soliton curvature 
radiation spectrum can be found by substituting $\mathcal{F}$ in 
eq.(\ref{eq:sol_spec}) with the relevant quantities in eq.(\ref{eq:curv_spec1})
-- (\ref{eq:curv_spec4}). Fig \ref{fig:sol_prop} panels (a) to (d), shows the 
changes of I, L, V and PPA, respectively, with the normalized angles 
$\gamma_o\theta$ and $\gamma_o\phi$. The Stokes parameters are estimated at the
characteristic frequency, $\omega \sim \omega_o$.

In the pulsar plasma the two perpendicular modes are separated and propagate
independently. If they remain unaltered during propagation then the spectral
distributions can be represented as :
\begin{eqnarray}
I_\parallel\left(\frac{\omega}{\omega_o},\theta,\phi, \gamma_o\right) & = & \gamma_o^2 \left(\frac{\omega}{\omega_o}\right)^2~(1 + \gamma_o^2\theta^2 + \gamma_o^2\phi^2)^2~\left[K_{2/3}^2(\xi) + \frac{\gamma_o^2\theta^2}{1 + \gamma_o^2\theta^2 + \gamma_o^2\phi^2}K_{1/3}^2(\xi)\right], \\
I_\perp\left(\frac{\omega}{\omega_o},\theta,\phi, \gamma_o\right) & = & \gamma_o^2 \left(\frac{\omega}{\omega_o}\right)^2~(1 + \gamma_o^2\theta^2 + \gamma_o^2\phi^2)(\gamma_o^2\phi^2)K_{1/3}^2(\xi),
\end{eqnarray}
The variations of $I_\parallel$ and $I_\perp$ with the two normalized angular
co-ordinates at the characteristic frequency, $\omega/\omega_o\sim1$, are shown 
in panels (e) and (f) of Fig \ref{fig:sol_prop}. The peak intensity of the 
parallel component is around seven times higher than the perpendicular 
component, which shows a double peaked structure along the $\phi$ co-ordinate. 
We are primarily interested in the average emission properties in this work and
will restrict our investigations to the vacuum like propagation in the 
remainder of the text. The parallel and perpendicular intensities will be 
useful when time varying properties are under consideration.

\subsection{Distribution of Solitons in Emission Region}

The radio emission region in pulsars is characterised by dipolar magnetic 
fields where neither the non-dipolar fields dominant near the surface 
\citep{BMM20b}, nor the sweep-back effects present near the light cylinder 
\citep{P16}, have any significant influence. For the calculations in this work 
we use a spherical co-ordinate system, ($r$, $\theta$, $\phi$), where the 
$z$-axis is aligned with the rotation axis. The magnetic field, $\bm{B} = (B_r, B_{\theta}, B_{\phi})$, due to a magnetic dipole, $\bm{d}$, inclined at an 
angle $\alpha$ with the rotation axis can be specified as : 
\begin{equation} \label{eq:magdip}
 B_r = \frac{2d}{r^3}(\sin{\alpha}\sin{\theta}\cos{\phi} + \cos{\alpha}\cos{\theta}),~~B_{\theta} = -\frac{d}{r^3}(\sin{\alpha}\cos{\theta}\cos{\phi} - \cos{\alpha}\sin{\theta}),~~B_{\phi} = \frac{d}{r^3}\sin{\alpha}\sin{\phi},
\end{equation}
where, $d\sim-10^{20}(P \dot{P}_{-15})^{-1/2}$ A$\cdot$m$^2$, can be estimated
from the measured pulsar parameters. The outflowing plasma is generated due to 
sparking discharges above the polar cap. The angular diameter of the dipolar
polar cap is estimated as $\theta_{PC} = 2\sin^{-1}\sqrt{R_S/R_{LC}}$, where, 
$R_S=10$ km. The solitons are formed in the secondary plasma moving along 
the open dipolar field line region, and their distribution is determined by the
plasma density which follows the particle density profile in the sparks. The 
range of heights over which the solitons are formed is further dependent on 
whether the conditions for the development of the non-linear plasma instability
is satisfied. We use the simplified dipolar geometry to populate the sparks 
across the open field lines and consider stationary behaviour without any 
drifting. The non-dipolar fields are essential for the formation of the sparks 
and they smoothly connect with the dipolar fields in the emission region 
\citep[see][for a more detailed discussion]{BMM22b,BMM23a}. In this work we are
primarily interested in the average features in the dipolar emission region, 
where the lack of non-dipolar magnetic fields and drifting in the spark 
configuration may remove certain asymmetric behaviour seen across the pulsar 
emission window, that will be explored in future works.

We initially use a primed co-ordinate system ($r$, $\theta'$, $\phi'$) where 
the $z'$-axis is aligned with the magnetic axis. Studies of the average 
emission beam \citep{ET_R93,MD99} suggest the presence of a central core 
component surrounded by either one or two conal rings, which we adapt for the 
arrangement of sparks above the polar cap. In the case of a single conal ring 
the diameter of each spark is $\theta_{sp}^{1c} = \theta_{PC}/3$ and the number
of sparks in the ring is $N_{1c} = 6$. The center of the $i^{th}$ spark on the
conal ring is specified by the angular separation from the axis $\theta'_{1c} =
\theta_{PC}/3$ and azimuthal angle $\phi'_i = 2\pi i/N_{1c}$, $i = 1, 2, ..., 
N_{1c}$. When two conal rings are present the angular width of each spark in 
$\theta_{sp}^{2c} = \theta_{PC}/5$. The centers of the inner and outer rings 
are inclined from the axis at angles $\theta'_{in} = 2\theta_{PC}/5$ and 
$\theta'_{out} = 4\theta_{PC}/5$, respectively, and the number of sparks in 
each ring are $N_{in} = 6$ and $N_{out} = 12$. The central location of $i^{th}$
spark on the inner and outer rings are specified by the azimuthal angles 
$\phi'_{in,i} = 2\pi i/N_{in}$, $i = 1, 2, ..., N_{in}$, and $\phi'_{out,i} = 
2\pi i/N_{out}$, $i = 1, 2, ..., N_{out}$. In addition, a central spark 
corresponding to the core component is located at $\theta'_{core}=\phi'_{core}=
0\degr$ in both configurations. At any height, $r$, the center of the sparks 
will follow the dipolar magnetic field lines and the angular separation from 
the axis will be 
\begin{equation}
\theta'_c(r) = \sin^{-1}\left(\sqrt{\frac{r}{R_S}}\sin{\theta'(R_S)}\right).
\end{equation}
The azimuthal angle remains unchanged, i.e. $\phi'_c(r) = \phi'_c(R_S)$, during
the transition, while the angular widths of the components at the emission 
height is obtained as $\theta_{cp} \sim \sqrt{r/R_S}~\theta_{sp}$. 

\begin{figure}
\gridline{\fig{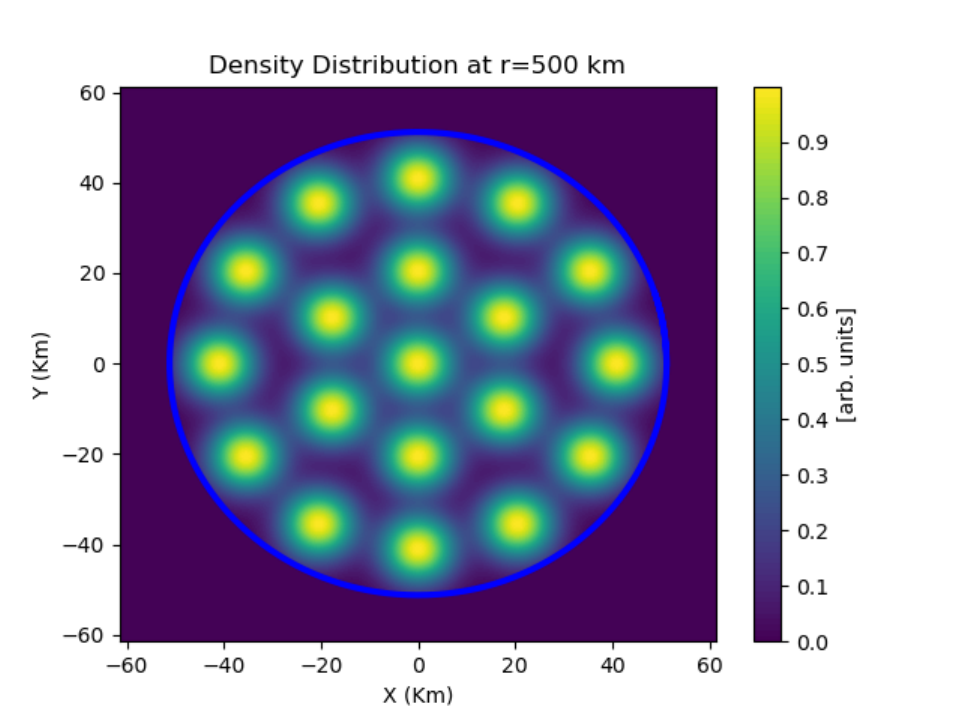}{0.52\textwidth}{(a)}
          \fig{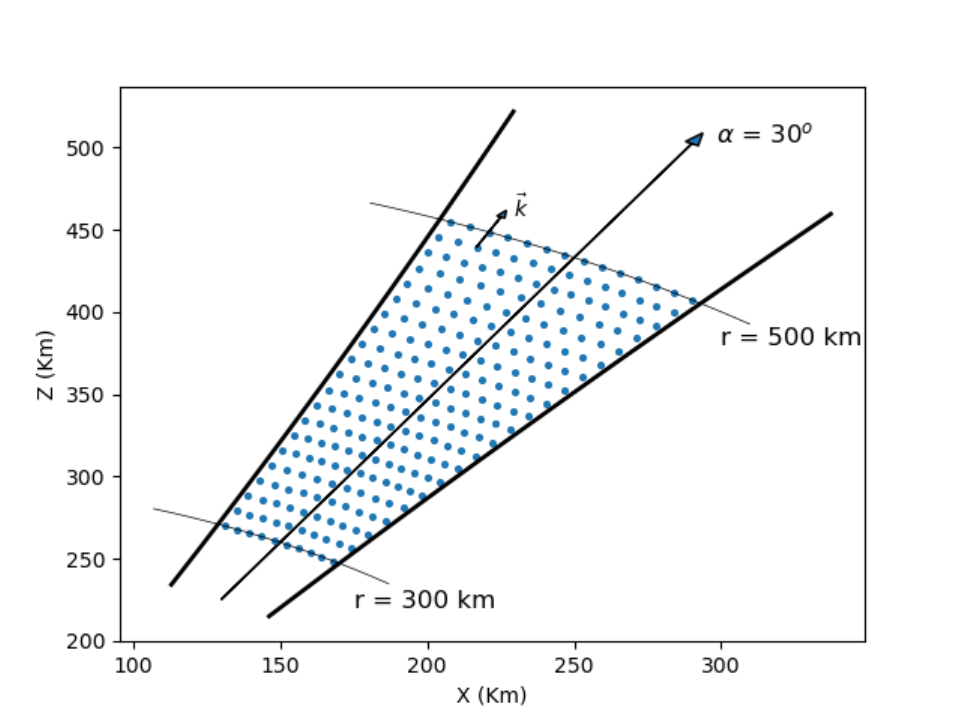}{0.52\textwidth}{(b)}
         }
\caption{The distribution of solitons along the open field line region for a 
pulsar with $P=1$ seconds and $R_{LC} = 4.77\times10^4$ km. (a) The 
density of the secondary plasma at a height of 500 km, obtained from the spark 
distribution above the polar cap. (b) The solitons with uniform distribution 
are located along the open field lines between heights of 300 km and 500 km. 
The inclination angle of the magnetic axis is $\alpha=30\degr$ and the 
direction of propagation of the radio emission, $\bm{k}$, is shown in one case.
\label{fig:sol_dist}}
\end{figure}

The density distribution of the secondary plasma is expected to be maximum at 
the center of the sparking column and decrease towards the boundary 
\citep{BMM22b,BMM23a}. We assume a Gaussian distribution of the density of the 
plasma columns such that :
\begin{equation}
\rho_{s} (x', y') = \rho_0(r) \exp{\left[-\frac{1}{2}\left((x' - x'_c)^2 + (y' - y'_c)^2\right)/a_{cp}^2\right]},
\end{equation}
where $(x', y')$ follows from co-ordinate transformation of $(r, \theta', 
\phi')$, $(x'_c, y'_c)$ represents the centers of the sparking columns 
specified by $(r, \theta'_c, \phi'_c)$, $\rho_0(r) \sim \kappa \rho_{GJ}(r) 
\propto r^{-3}$ and $a_{cp} \sim r\theta_{cp}/2$. Finally, the entire system is
rotated by an angle $\alpha$ around the $y$-axis to align with the pulsar. Fig 
\ref{fig:sol_dist}(a) shows an example of the horizontal cross section of the 
secondary plasma density distribution within the open field line boundary at 
$r=500$ km for a system with $P=$ 1 seconds, such that $R_{LC} = 
4.77\times10^4$ km, and two conal rings.

The conditions for formation of solitons and the emergence of radio emission 
is satisfied within heights of around 100 -- 1000 km \citep[see Discussion 
in][]{MMB23b}. Fig \ref{fig:sol_prop} shows that the intensity of curvature 
radiation from a soliton is confined within an angular scale specified by 
$1/\gamma_o \sim 1/y\gamma_s$. A distribution of solitons with angular 
separation $1/2y\gamma_s$ in the open field line region is sufficient to obtain
a continuous intensity pattern from pulsars without any jumps or gaps. At any 
given frequency the radio emission due to CCR from solitons is generated within
a narrow range of heights with upper and lower limits $r_L$ and $r_H$, 
respectively. To simplify the calculations we divide this region into multiple 
grids of radial size $\delta r$$\sim$0.1--1$R_S$ (see discussion below) and 
angular size $\delta\theta = \delta\phi = 1/2y\gamma_s$. The solitons are 
located at the center of each grid and emit curvature radiation along the 
direction of propagation which is tangent to the magnetic field line at the 
grid center. The contribution of all solitons along a longitudinal phase is 
added up to obtain the emission pattern of the specific pulsar configuration as
detailed in the next section. At an emission height $r=100$ km, angular 
resolution specified by $\gamma_s = 100$ and $y = 2.3$, the horizontal 
separation between two adjacent grids is $r\delta\theta = r/2y\gamma_s \sim 
200$ m. Using typical wavelength of radio waves, $\lambda\sim 1$ m, the maximum
soliton size is $d \lesssim y \gamma_s \lambda/2 \sim 100$ m, which makes it 
physically viable to fit the solitons within the grid cross section. The 
density of plasma at the location of solitons is specified by $\rho_{s}$, such 
that charge of each soliton has the dependence $Q_s \propto \rho_s r^2 
\Delta_s/\gamma_s^2$. In the radial direction more solitons can be filled up 
within the grid whose number is obtained as $N_s=\delta r/\Delta_s$. Fig 
\ref{fig:sol_dist}(b) shows a schematic of the soliton distribution for a 
pulsar with inclination angle $\alpha = 30\degr$, where the emission arises 
between $r_L = 300$ km and $r_H = 500$ km, while Fig \ref{fig:sol_avg}(b) shows
the variation in plasma density across the open field region at $r = 400$ km.

\section{Average Emission Features from Solitons} \label{sec:Avgem}

\begin{figure}
\gridline{\fig{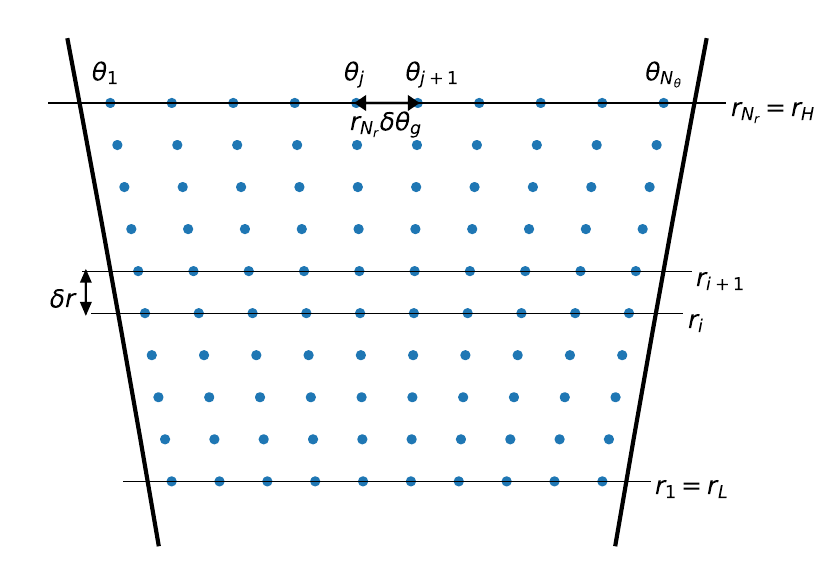}{0.52\textwidth}{(a)}
          \fig{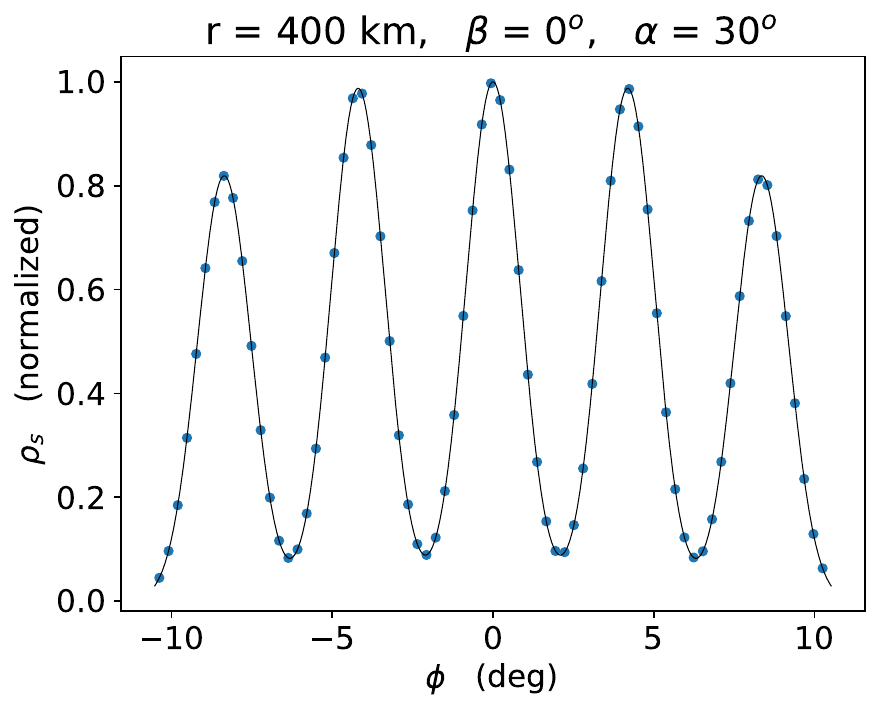}{0.42\textwidth}{(b)}
         }
\caption{(a) A schematic showing the arrangement of solitons in the open field 
line region used for estimating the average emission properties from different 
pulsar configurations. The emission region is split into small grids with 
solitons located at the center (blue dots) of each grid, $\bm r_g = (r_i, 
\theta_j, \phi_m)$, and has a three dimensional size $\Delta V = (\delta r, 
r_i\delta\theta_g, r_i\delta\theta_g)$ (see text for details). The plot shows 
a two dimensional slice of the grids along the $r$ and $\theta$ co-ordinates. 
(b) The plasma density, $\rho_{s}$, at the location of the solitons (blue dots)
as a function of the angular co-ordinate $\phi$ at a distance of $r=400$ km and
$\theta = \alpha$, i.e. LOS traversing centrally across the magnetic axis with
$\beta =0\degr$. The sparking pattern at the surface of the polar cap comprises
of a central spark surrounded by two concentric rings of inner and outer sparks
(see Fig \ref{fig:sol_dist}a), resulting in the five distinct plasma columns 
seen in the plot.
\label{fig:sol_avg}}
\end{figure}

The emission from solitons in the pulsar plasma can be obtained from the setup 
specified above, by suitably adding their contributions along a given 
direction. This also involves specifying range of several physical parameters 
that determine the emission from the plasma which are listed below. We have 
further assumed that once the plasma waves are excited by the solitons they 
have vacuum like propagation properties and are not further modified within the
magnetosphere.
\begin{enumerate}
\item {\it Distribution of secondary plasma} : We are interested in estimating 
the average emission properties from pulsars over long timescales to compare 
with the observational features. The secondary plasma formed as a result of the
sparking process in the IAR have a range of energies with Lorentz factors 
$\gamma_s\sim100$ \citep{S71,TH19}. This implies that at any particular instant
the $\gamma_s$ at each grid center indeed has an unique value, but at a 
different instance the $\gamma_s$ is likely to change. If long enough durations
are considered then the average emission from each grid should have 
contribution from the entire range of $\gamma_s$ specified by their energy
distribution, which we assume to be uniform between $\gamma_l$ = 50 and 
$\gamma_h$ = 300, unless otherwise stated, with a step size of $\gamma_{\rm 
step}$ = 10. The multiplicative factor used for the soliton group velocity is 
$y=2.3$. An average angular resolution for each grid is obtained as $\delta 
\theta_g = 1/y(\gamma_l+\gamma_h)$.
\item {\it Characteristic length of solitons} : The soliton length varies as a
function of $\gamma_s$ and the altitude, $r$, as well as the strength of the 
plasma instability leading to charge bunching specified by characteristic 
length $\Delta_o$ in eq.(\ref{eq:sol_len}) \citep{MGP00}. At $r=500$~km and 
$\gamma_s=100$ the lower limit of the length has been set as $\Delta_l$ = 0.4~m
and upper limit as $\Delta_h$ = 0.6~m, with a step size of 0.1~m.
\item {\it Emission heights} : Depending on the pulsar parameters and the 
frequency range under consideration, there is variation in the range of heights
over which the radio emission emerge (see section \ref{sec:RFM}). The range of 
heights and the radial size of each grid are estimated in an iterative manner, 
where initially we assume $r_L$ = 50 km, $r_H$ = 1000 km and $\delta r$ = 10 km 
and estimate the emission from each grid. Subsequently, the emission height 
range, $r_L$ and $r_H$, is restricted to the narrower window with significant 
levels of emission, i.e. $I_g > 0.001 I_{\rm max}$. The radial size of each 
grid is recalculated as $\delta r = (r_H-r_L)/100$ to ensure uniform 
contribution to the average emission features from different heights within 
this range.
\item {\it Frequency range} : Pulsars are primarily observed at frequency
range between 100 MHz and 1.4 GHz, with 150~MHz, 325~MHz, 610~MHz and 1.4~GHz 
being prominent frequencies for pulsar studies. The emission properties are 
estimated between lower frequency range $\nu_l$ = 100 MHz and upper level 
of $\nu_h$ = 2000 MHz, with multiplicative increment in step size. 
\end{enumerate}

Several detailed calculations are carried out to simulate the pulsar emission 
from solitons. These are implemented using specific codes developed utilizing 
the standard {\it scipy} package of {\it python}. A series of steps are listed 
below describing the implementation plan.
\begin{enumerate}
\item The first step involves specifying the central position of each grid, 
$\bm r_g = (r_i, \theta_j, \phi_m)$, where the integers $i$, $j$ and $m$ 
uniquely identifies each grid as explained below. Fig. \ref{fig:sol_avg}(a) 
shows schematic of two dimensional projection of the grid layout along the $r$
and $\theta$ co-ordinates. 
\begin{enumerate}
\item The radial position of the grid is specified as, $r_i$ = $r_L$ + $(i - 
1/2)\delta r$, where $i = 1, 2, ..., N_r$ and $N_r = 100$, such that $r_H = r_L
+ N_r\delta r$ (see Fig. \ref{fig:sol_avg}a).
\item At $r_i$ the maximum opening angle of the emission beam is :
\begin{equation}
\theta^i_{\rm bm} = \sin^{-1}\sqrt{r_i/R_{LC}}.
\label{eq:th_grid}
\end{equation}
When the inclination angle between the rotation and magnetic axis is $\alpha$, 
the $\theta$ co-ordinate varies between $\alpha - \theta^i_{\rm bm}$ and 
$\alpha + \theta^i_{\rm bm}$, such that the $j^{th}$ grid can be identified as 
$\theta_j = \alpha - \theta^i_{\rm bm} + (j-1/2)\delta\theta_g$, where $j = 1, 
2, ..., N^i_{\theta}$ and $N^i_{\theta} = 2\theta^i_{\rm bm}/\delta\theta_g$. 
The length of the grid along the $\theta$ co-ordinate is $\delta {r_{\theta}} =
r_i\delta\theta_g$ (see Fig. \ref{fig:sol_avg}a).
\item The maximum azimuthal angle, $\phi^{ij}_{\rm max}$, across the 
emission beam for the co-ordinates $r_i$ and $\theta_j$ can be estimated using 
spherical geometry \citep{GGR84}. An expression for $\phi^{ij}_{\rm max}$ in
terms of the inclination angle, $\alpha$, and the beam opening angle, 
$\theta^i_{\rm bm}$, has the form \citep[see eq.2 in][]{ET_R93} :
\begin{equation}
\phi^{ij}_{\rm max} = 2 \sin^{-1}\left[\cfrac{\sin{(\theta^i_{\rm bm}/2 + \theta_j/2)}\sin{(\theta^i_{\rm bm}/2 - \theta_j/2)}}{\sin{\alpha}\sin{(\alpha+\theta_j)}} \right]^{1/2} 
\label{eq:phigrid}
\end{equation}
The azimuthal angle varies between -$\phi^{ij}_{\rm max}$ and $\phi^{ij}_{\rm 
max}$, and the center of the $m^{th}$ grid is identified as $\phi_m = 
-\phi^{ij}_{\rm max} + (m-1/2)\delta\theta_g$, $m = 1, 2, ..., N^{ij}_{\phi}$ 
and $N^{ij}_{\phi} = 2\phi^{ij}_{\rm max}/\delta\theta_g$.
\end{enumerate}

\item We next estimate the propagation direction, $\bm{k_g} = (\theta_k, 
\phi_k)$, of tangential direction at center of each grid located at $\bm r_g$. 
The propagation direction is determined by the tangent along the magnetic field
line and the co-rotation velocity, and can be expressed as \citep{DRH04,D08} :
\begin{equation}
\bm{k_g} = \frac{\bm{k'} + [\gamma_{rot} + (\gamma_{rot}-1)(\bm{\beta}_{rot}\cdot\bm{k'})/\beta_{rot}^2]\bm{\beta}_{rot}}{\gamma_{rot}(1 + \bm{\beta}_{rot}\cdot\bm{k'})}.
\label{eq:kvec}
\end{equation}
Here, $\bm{k'} = \bm{B}(\bm{r_g})/B$, $\bm{\beta}_{rot} = \bm{\Omega} \times 
\bm{r_g}/c$ and $\gamma_{rot} = 1/\sqrt{1-\beta_{rot}^2}$.

\item The direction of the electric vector along the center of the grid, 
$\bm r_g$, is rotated by an angle $\chi_g$ with respect to the plane containing
the magnetic axis and the rotation axis which has the dependence \citep{BCW91,
D08} :
\begin{equation}
\chi_g = \tan^{-1}\left[\cfrac{-\sin{\alpha}\sin{(\phi_m+r_i/R_{LC})} + 3(r_i/R_{LC})\sin{(\alpha+\theta_j)}}{\sin{(\alpha+\theta_j)}\cos{\alpha} - \cos{(\alpha+\theta_j)}\sin{\alpha}\cos{(\phi_m+r_i/R_{LC})}} \right].
\label{eq:PPA_ht}
\end{equation}
The estimated shift is approximate but remains valid for low emission heights
$r_i/R_{LC}\lesssim0.1$, which is typically satisfied for longer period pulsars
with $P>0.1$ seconds.

\item The average emission properties are estimated along a line of sight (LOS)
specified by the impact angle $\beta$, which represents the angle between the 
magnetic axis and the LOS during closest approach. The angle between the 
rotation axis and LOS is obtained as $\theta^L = \alpha + \beta$, while the 
azimuthal angle is a variable specified by $\phi_l^L = -\pi + l\delta\theta_g$,
$l = 1, 2, ..., N_L$, and $N_L = 2\pi/\delta\theta_g$. The different emission 
properties are estimated using $\theta^L$ and $\phi_l^L$, where $\phi_l^L$ 
corresponds to the longitude variation.

\item The total emission is determined by adding the contribution of all 
solitons whose emission is directed towards the LOS specified by the unit 
vector $\bm{n_L} = (\theta^L, \phi_l^L)$, under the following constraints : 
\begin{enumerate}
\item The angular separation between the tangent vector at the center of every
grid and the specific LOS vector is estimated as : $\cos{\varphi_g} = 
\bm{k_g}\cdot\bm{n_L}/k_g$, and the grids where $\gamma_o\varphi_g\leq1.5$ can 
contribute sufficient intensities to the radio emission along the LOS (see Fig 
\ref{fig:sol_prop}).

\item In all the relevant grids the angular separation between the LOS and 
$\bm{k_g}$ is estimated along the angular co-ordinates, $\Delta\theta = 
\theta_k - \theta^L$ and $\Delta\phi = \phi_k - \phi_l^L$. The four Stokes 
parameters corresponding to each soliton towards the LOS can be estimated using
eq.(\ref{eq:curv_spec1}) -- (\ref{eq:curv_spec4}) by substituting the 
appropriate angles,  $I (\omega/\omega_o, \Delta\theta, \Delta\phi, \gamma_o)$,
$Q (\omega/\omega_o, \Delta\theta, \Delta\phi, \gamma_o)$, $U (\omega/\omega_o,
\Delta\theta, \Delta\phi, \gamma_o)$ and $V (\omega/\omega_o, \Delta\theta, 
\Delta\phi, \gamma_o)$.

\item Additional rotation of the Stokes parameters by angle $\chi_g$ is carried
out for precise alignment with the reference plane, before they are added. This
involves estimating $L = \sqrt{Q^2+U^2}$ and $\chi_i = 0.5\tan^{-1}(U/Q)$ to 
obtain the rotated quantities $Q_f = L \cos{2(\chi_i+\chi_g)}$ and $U_f = L 
\sin{2(\chi_i+\chi_g)}$.

\item The entire process is repeated for the range of parameter described 
earlier, including the heights between $r_L$ and $r_H$, the distribution of the
Lorentz factors of secondary plasma between $\gamma_l$ and $\gamma_h$, the 
soliton lengths between $\Delta_l$ and $\Delta_h$, across a frequency band, to 
obtain the total radio emission intensity along any LOS. 
\end{enumerate}

\end{enumerate}
In the next subsections we investigate how the radio emission features from 
solitons obtained from the above setup compares with some of the established
observational results in pulsars.

\subsection{Profile Types from Line of Sight Trajectory} \label{sec:profclass}

\begin{figure}
\gridline{\fig{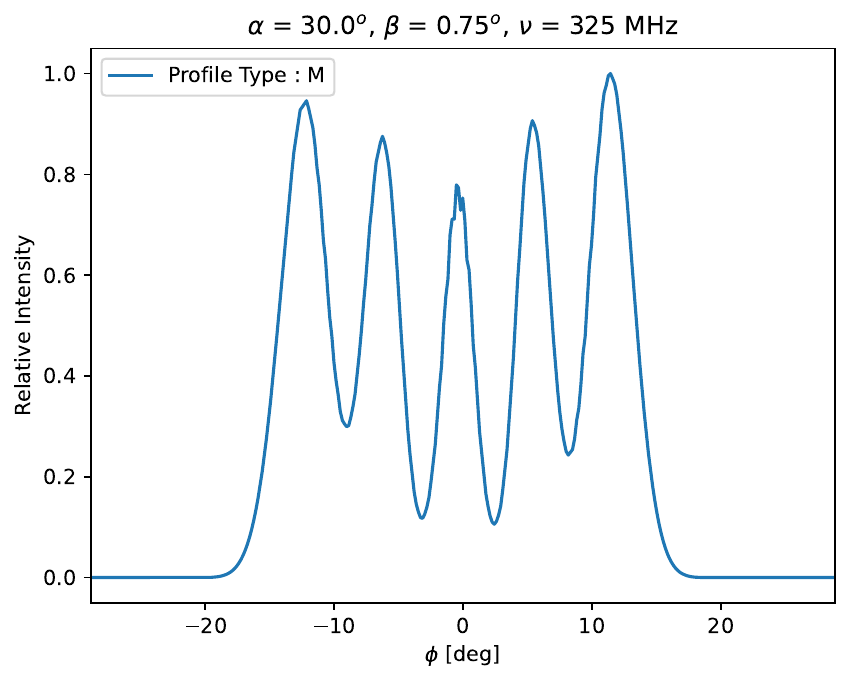}{0.42\textwidth}{}
          \fig{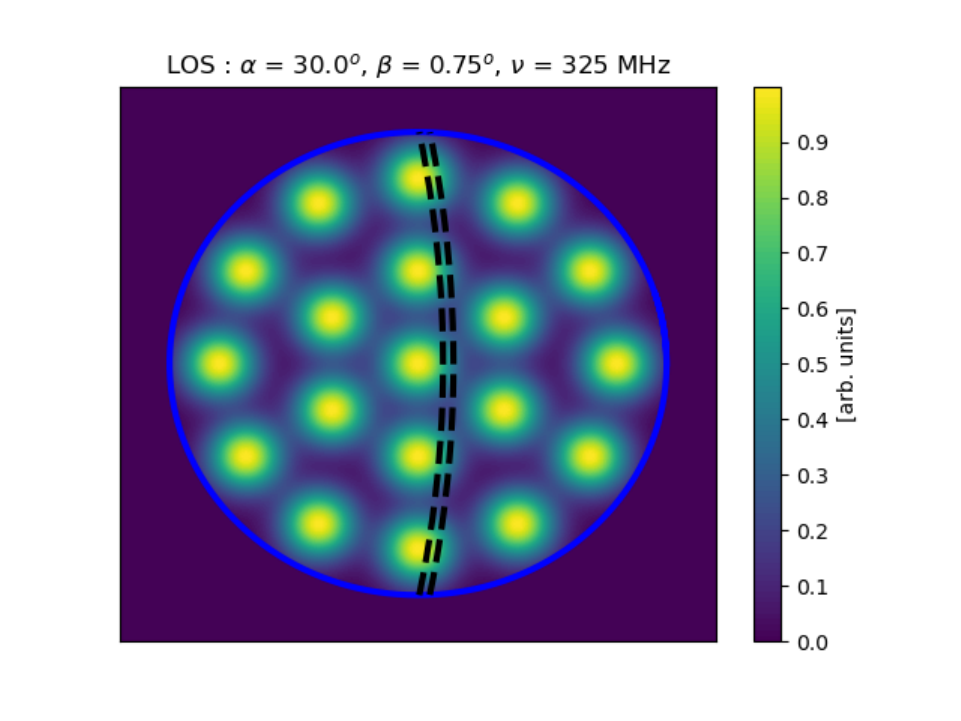}{0.48\textwidth}{}
         }
\gridline{\fig{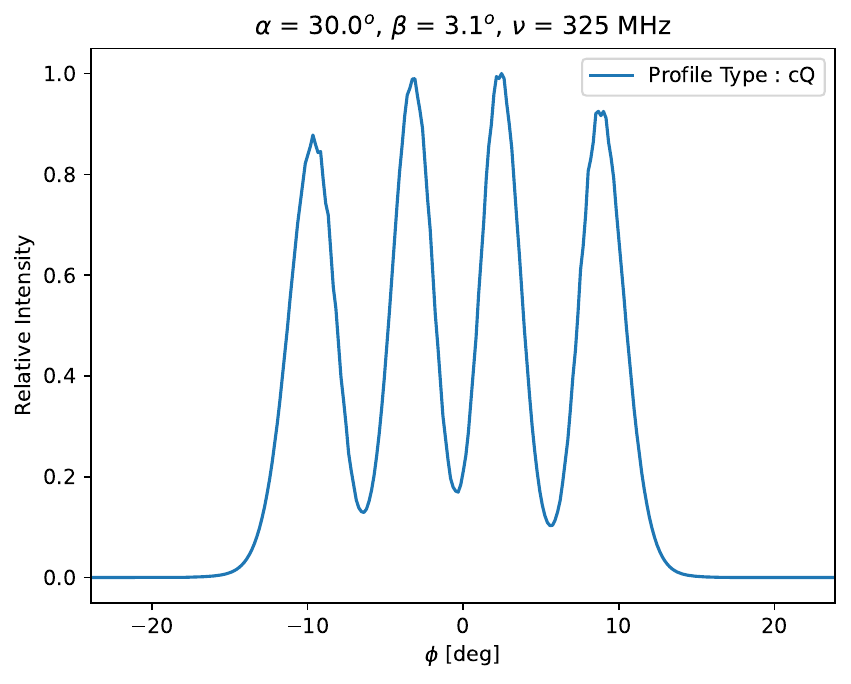}{0.42\textwidth}{}
          \fig{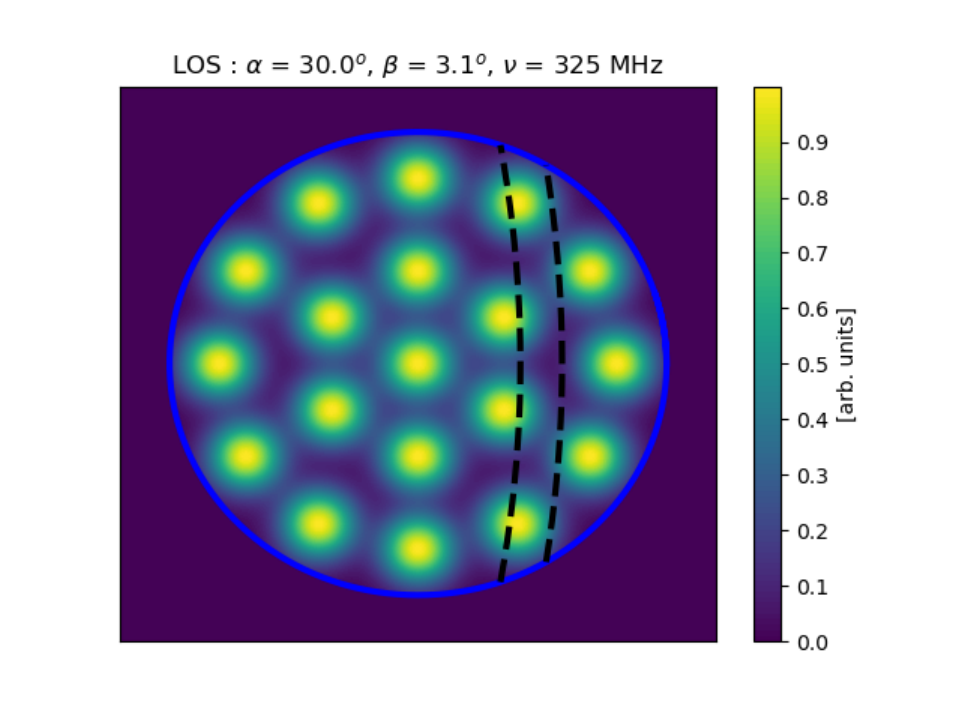}{0.48\textwidth}{}
         }
\caption{The top figures show the multiple (M) profile consisting of a central 
core and two pairs of inner and outer conal emission. The line of sight cuts 
close to the magnetic axis with impact angle $\beta=0.75\degr$. The bottom 
panels show the $_c$Q type profile, with four conal components, that appears 
when the line of sight cut across the emission region is further away from the 
magnetic axis with an impact angle $\beta=3.1\degr$. The two dotted lines in 
each figure on the right show the area of the emission beam covered by the LOS 
over the range of emission heights at 325 MHz (see section \ref{sec:RFM}).
\label{fig:prof2cone}}
\end{figure}

The average profiles in pulsars represent the intensity pattern along 
longitudes of the LOS, and comprise of up to five components that are unique to
each pulsar. Despite the large variability of profile types in the pulsar 
population there appears to be an underlying pattern in their shapes. The 
profiles have been classified using the core-cone model of the emission beam
\citep{GKS93,ET_R93}, where the radio emission emerges from a beaming pattern
from the open field line region, comprising of a central core region surrounded
by one or two concentric conal emission rings. When the LOS cuts the beam close
to the magnetic axis, i.e. small value of $\beta$, the core component is seen
surrounded by one or two conal pairs and form either triple (T) or multiple (M)
profile types, respectively. When the LOS traverse is further away from the  
magnetic axis with larger values of $\beta$, the core emission is missed and 
the profile has types conal Quadruple ($_c$Q), conal Triple ($_c$T), Double (D)
and conal single (S) profiles, with correspondingly larger values of $\beta$. 
In some cases the core single profiles are also seen when the LOS traverse 
passes through the central core and the surrounding cones are not visible. The
PPA variations are also different across the core and conal components. In the 
core component the PPA show large swing with a steep gradient, while they are 
relatively flatter across the cones.

\begin{figure}
\gridline{\fig{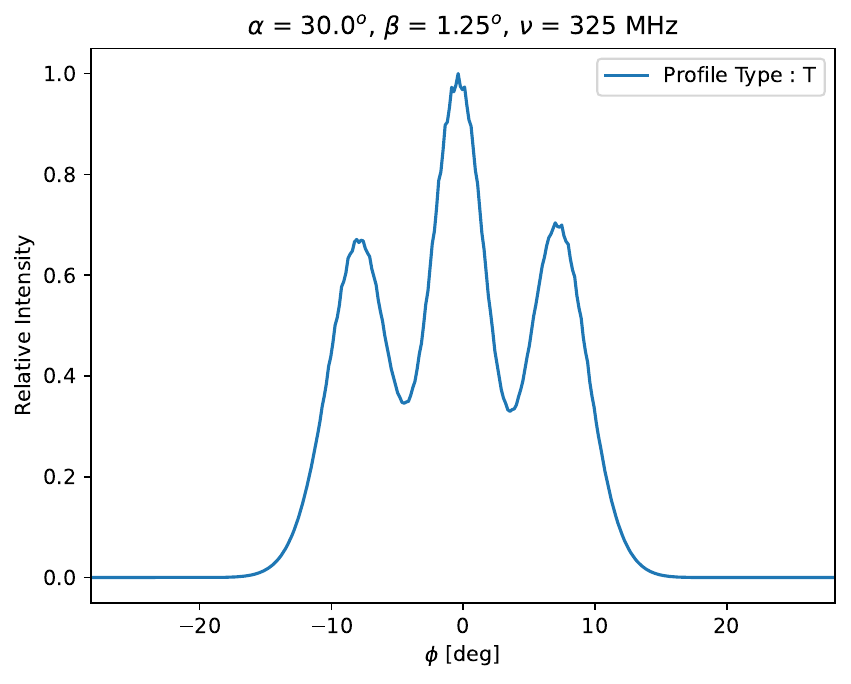}{0.4\textwidth}{}
          \fig{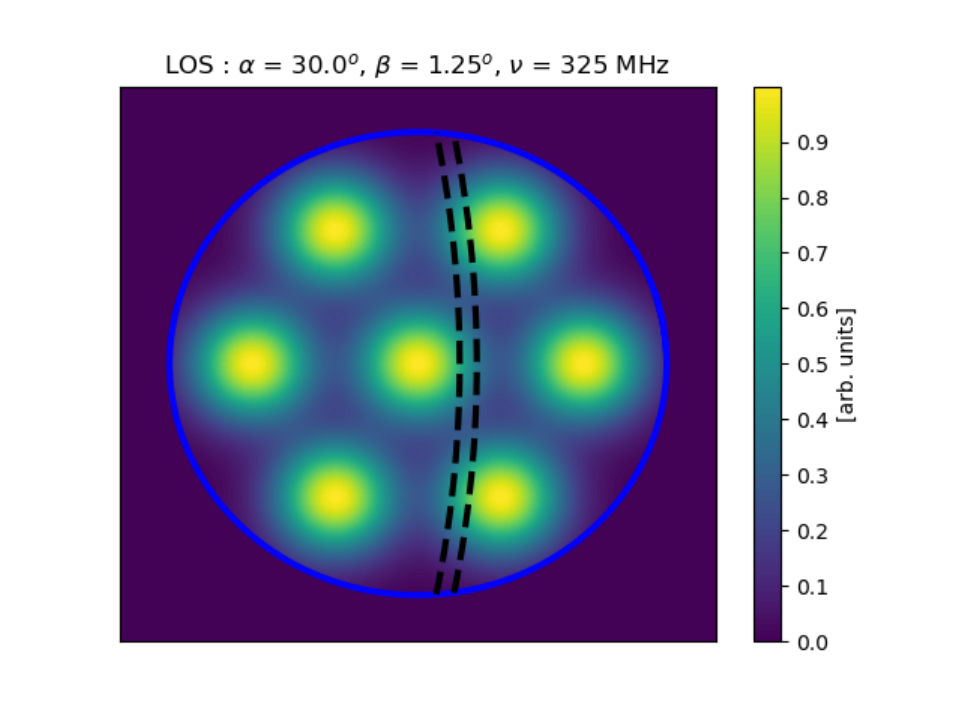}{0.46\textwidth}{}
         }
\gridline{\fig{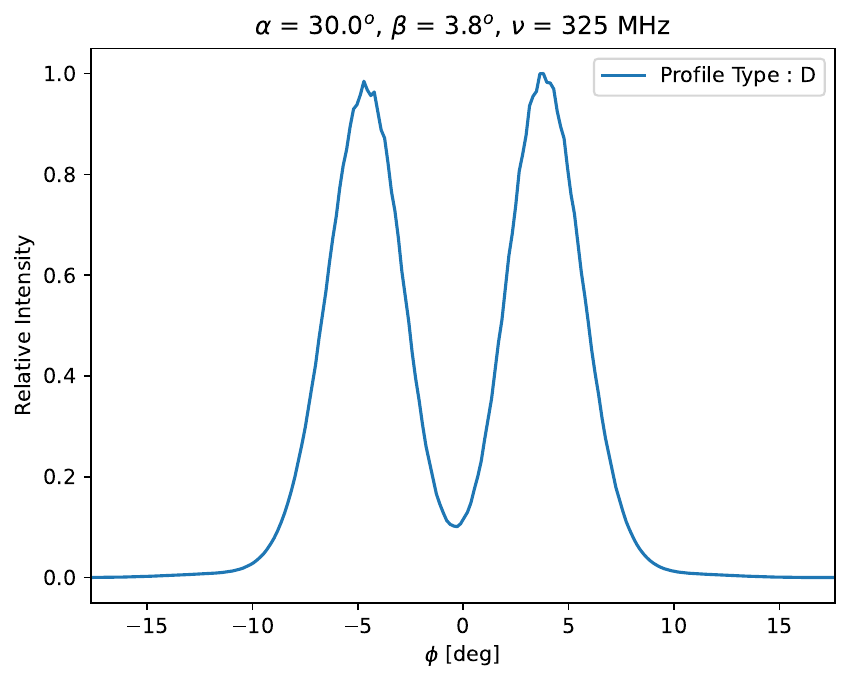}{0.4\textwidth}{}
          \fig{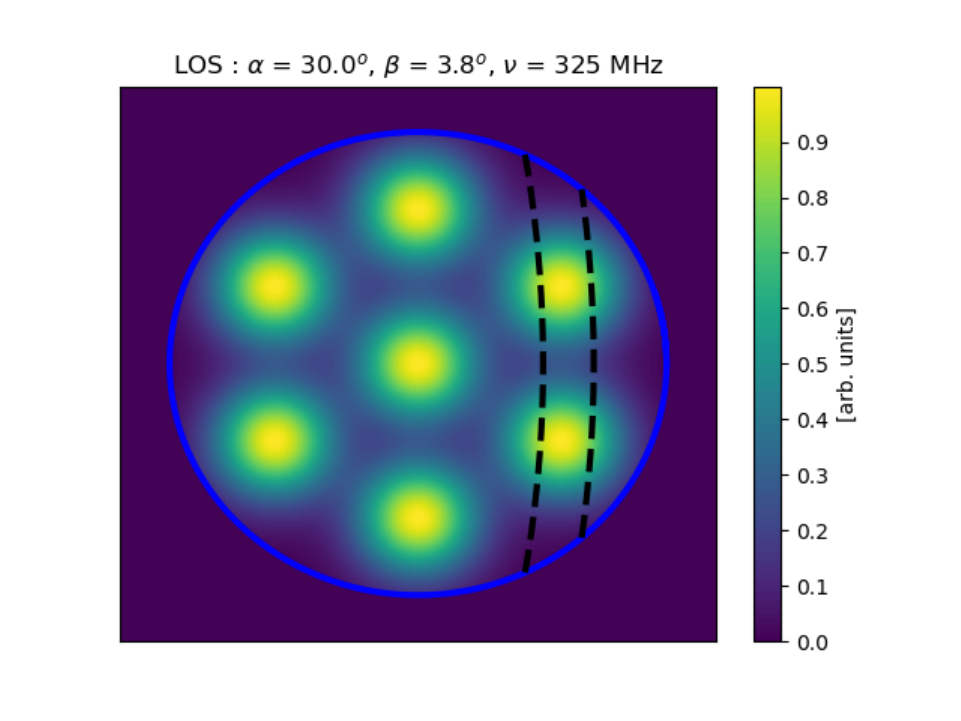}{0.46\textwidth}{}
         }
\gridline{\fig{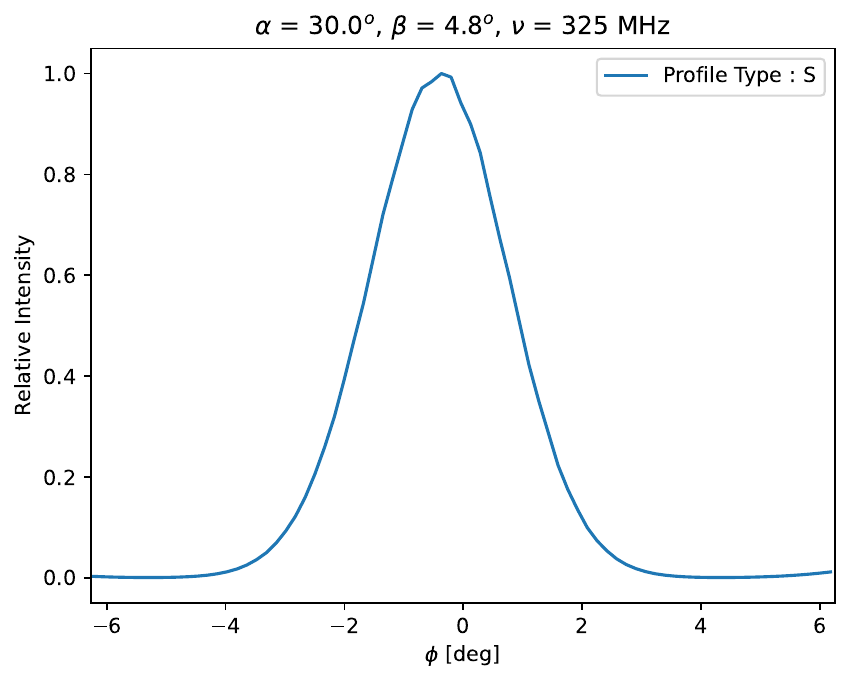}{0.4\textwidth}{}
          \fig{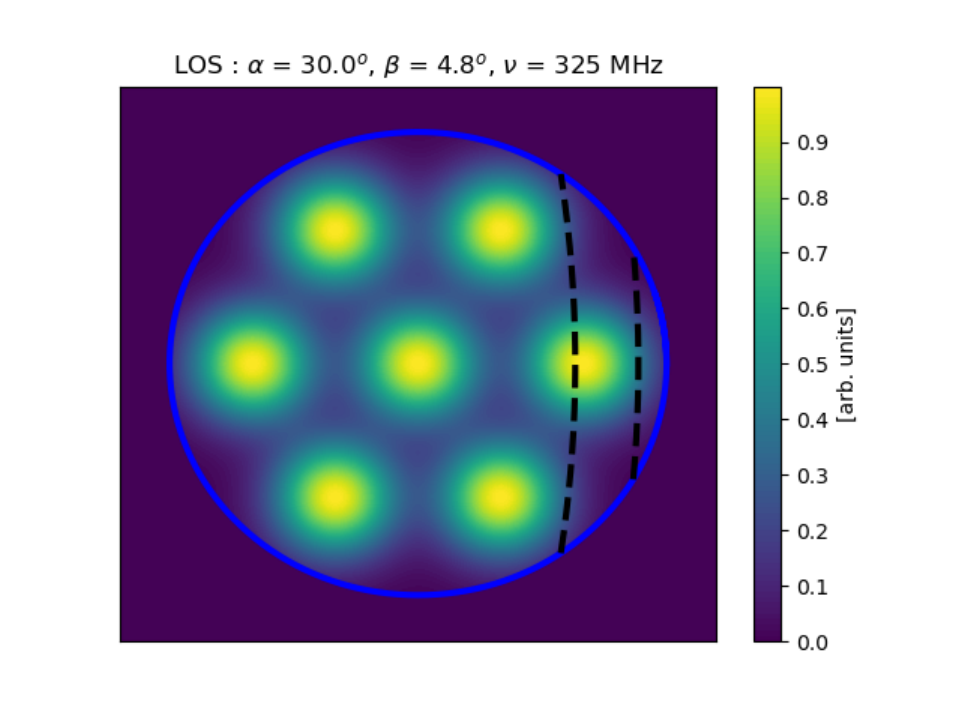}{0.46\textwidth}{}
         }
\caption{The figure shows the effect of different line of sight cuts across the
emission region originating from a sparking pattern with one conal ring around 
a central spark. The top panel corresponds to a core-cone triple (T) profile 
when the impact angle $\beta=1.25\degr$, the middle panel shows the double (D) 
profile when $\beta=3.8\degr$, while the bottom panel shows a conal single (S) 
profile when the line of sight cuts near the periphery with $\beta = 4.8\degr$.
The two dotted lines on the right panels show the area of the emission beam 
covered by the LOS over the range of emission heights at 325 MHz (see section 
\ref{sec:RFM}).
\label{fig:prof1cone}}
\end{figure}

Due to the difference in their properties it has been suggested that core 
emission has a different origin compared to the cones, with the core located
close to the surface and has a different emission mechanism than the cones 
\citep{ET_R90}. However, such drastic deviations are not needed if the core 
component arises from the plasma generated by the central spark, while the 
conal emission arises from the surrounding sparks. To study the effect of 
different LOS cuts we have considered two sparking configurations, one with two
rings around a central spark and the second with only one conal ring, and the 
effect of different LOS cuts across them is described below.
\begin{enumerate}
\item {\it Profile type M} : Fig \ref{fig:prof2cone} (top panel), shows one 
example of a M type profile estimated between $\nu_l = 300$ MHz and $\nu_h = 
350$ MHz, where the pulsar has inclination angle $\alpha = 30\degr$ and the 
central LOS is specified by $\beta = 0.75 \degr$. The outflowing plasma is 
generated from a distribution of two rings around a central spark. The radio 
emission at a given frequency arises from a range of heights (see section 
\ref{sec:RFM}) such that the LOS cuts across different regions of the the spark 
associated plasma columns at each height range. The right panel in the figure 
shows the horizontal cross section of the secondary plasma distribution where 
the area between the two dotted lines include LOS traverses due to all possible
emission heights. The five component profile shown in the left window is 
obtained as the LOS cuts the central spark as well as the two sparks from the 
inner and outer rings on either side.
\item {\it Profile type $_c$Q} : The bottom panel of Fig \ref{fig:prof2cone}
shows the $_c$Q profile obtained when the LOS traverse is further away from
the magnetic axis with $\beta = 3.1 \degr$. The spark configuration and the 
remaining parameters are identical to the previous case. The LOS traverse 
covers a wider cross section across the emission beam due to diverging nature 
of the magnetic field lines. The LOS misses the central spark, but cuts across
the two components of the inner and outer rings resulting in the four component
profile.
\item {\it Profile type T} : Fig \ref{fig:prof1cone} (top panel) shows a triple
profile obtained from a spark pattern distributed in one ring surrounding a 
central spark. The LOS is close to the center with $\beta = 1.25 \degr$, (see 
right window) and passes through the central core component as well as the pair
of conal components, seen as the three component profile in the left window.
\item {\it Profile type D} : The plasma distribution in this case is modified
compared to the triple profile case, with the sparks rotated by 30$\degr$ in
the conal ring. An outer LOS traverse with $\beta = 3.8 \degr$ passes through
two sparks of the conal ring, bypassing the central spark (see right window in
the middle panel of Fig \ref{fig:prof1cone}). The resultant profile has a 
double peaked structure as shown in the left window.
\item {\it Profile type S} : The bottom panel in Fig \ref{fig:prof1cone} shows 
a single component profile obtained from a plasma distribution identical to the
triple profile, but a LOS located towards the edge of the open field line 
region with $\beta = 4.8 \degr$ (see right window). Only a single conal 
component is visible in the average profile as seen in the left window of the
panel. It is also possible to obtain a single component profile when the LOS 
passes through the center of the emission window. In such cases the core single
profile can be seen with the conal emission being weak or absent as the LOS 
cuts the low density region between adjacent conal plasma columns.
\end{enumerate}

The above exercise is intended to show the effect of the LOS geometry on the 
profile types and demonstrate that the wide variety of profiles seen in pulsars
can be explained by solitons emerging from the spark associated secondary 
plasma. The asymmetry seen in pulsar profiles can also be modelled by 
considering the non-dipolar fields above the surface where the sparks are 
formed and connect with the dipolar emission region. The surface magnetic field
variations \citep[see][]{GBM21} can further introduce density difference across
the open field lines resulting in suppression or enhancement of the emission at
different longitudes of the profile. In all estimates above we have considered
the pulsar period to be 1 second. For longer periods the opening angle of the
emission beam become smaller and it is possible to have a single spark in long 
period pulsars like PSR J2144$-$3933 \citep{MBM20}. On the other hand faster 
pulsars will have much wider open field line regions and can accommodate a 
larger number of sparks. The close packing of the emergent emission components 
can blur the separation between them resulting in merged profiles \citep{GS00,
MMB23b}.

\subsection{Frequency Dependence of Emission} \label{sec:RFM}

\begin{figure}
\gridline{\fig{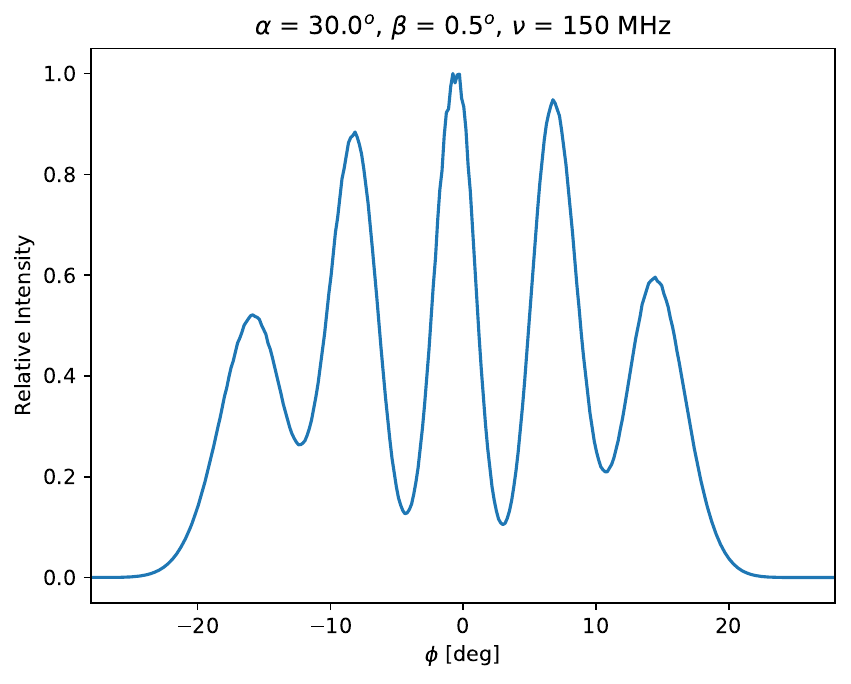}{0.42\textwidth}{}
         \fig{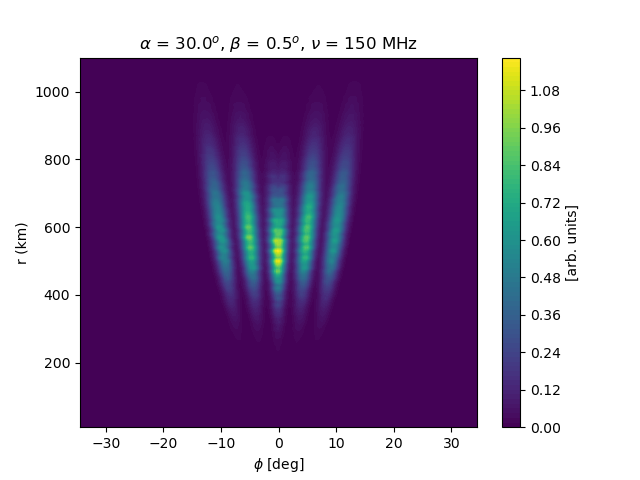}{0.48\textwidth}{}
         }
\gridline{\fig{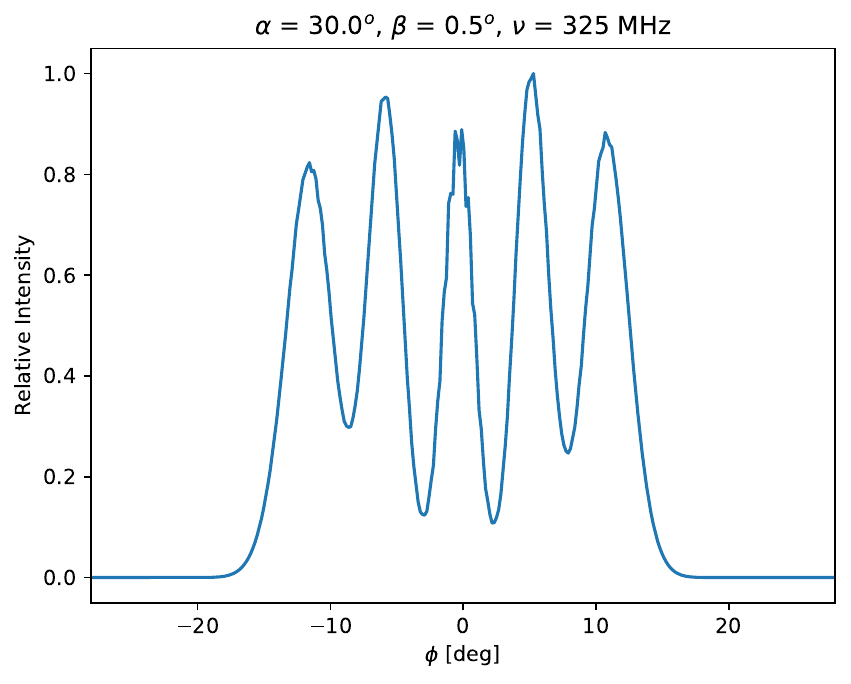}{0.42\textwidth}{}
          \fig{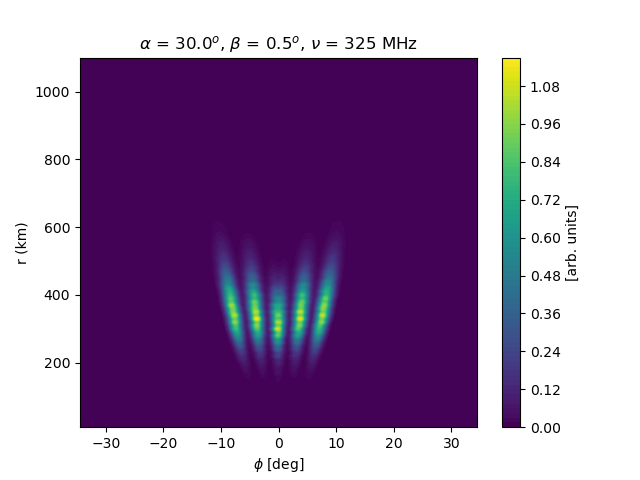}{0.48\textwidth}{}
         }
\gridline{\fig{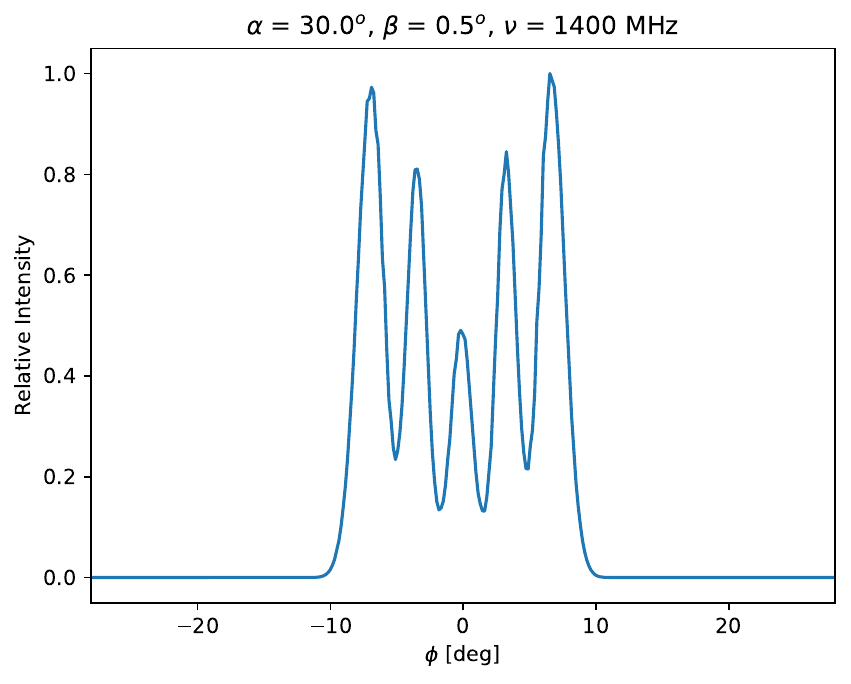}{0.42\textwidth}{}
          \fig{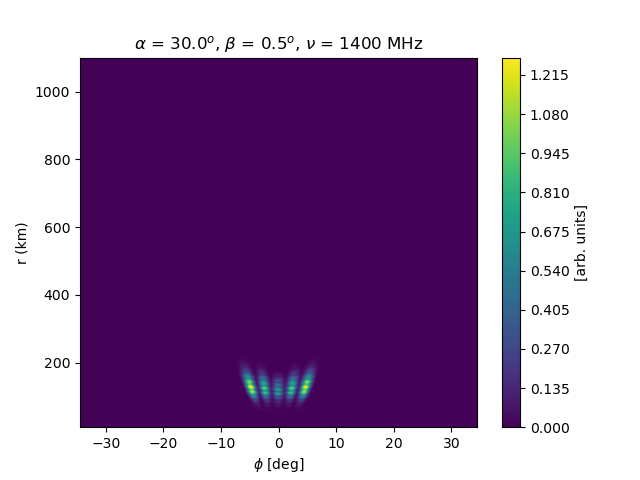}{0.48\textwidth}{}
         }
\caption{The figure shows the average profiles at three frequencies, 150 MHz, 
325 MHz and 1400 MHz for the same line of sight traverse across emission region
along with the locus of the radio emission heights.
\label{fig:RFM_ht}}
\end{figure}

The frequency dependence of the emergent radio emission from CCR due to charged
solitons arises from two primary constraints :
\begin{enumerate}
\item The characteristic frequency is dependent on the radius of curvature of 
the magnetic field line, $\omega_o \propto 1/\mathcal{R}$, where $\mathcal{R}$ 
changes both as function of LOS traverse, $\phi_l^L$, and the emission height, 
$r$ (see Fig \ref{fig:rad_curv} in Appendix \ref{app:radcurv}).
\item The coherence condition for the curvature radiation from solitons 
requires the soliton size to be less than the half-wavelength, i.e. $\Delta_s <
\lambda/2$. The soliton size is dependent on the emission height (see 
eq.\ref{eq:sol_len}), such that emission height has an upper limit with a 
frequency dependence, $r_U \propto \nu^{-2/3}$.
\end{enumerate}
To study the frequency dependence of the radio emission we consider a central
LOS with $\beta = 0.5\degr$ for a pulsar with $P=1$ seconds and $\alpha = 
30\degr$, which results in a M type profile where both the core and conal 
components are seen. The range of heights over which the radio emission emerges
are estimated by adding up the total contribution of the emission from each 
grid along the open field line and generating a two-dimensional contour plot of
grid intensities with respect to $\phi_l^L$ and $r$. 

\begin{figure}
\epsscale{1.1}
\plottwo{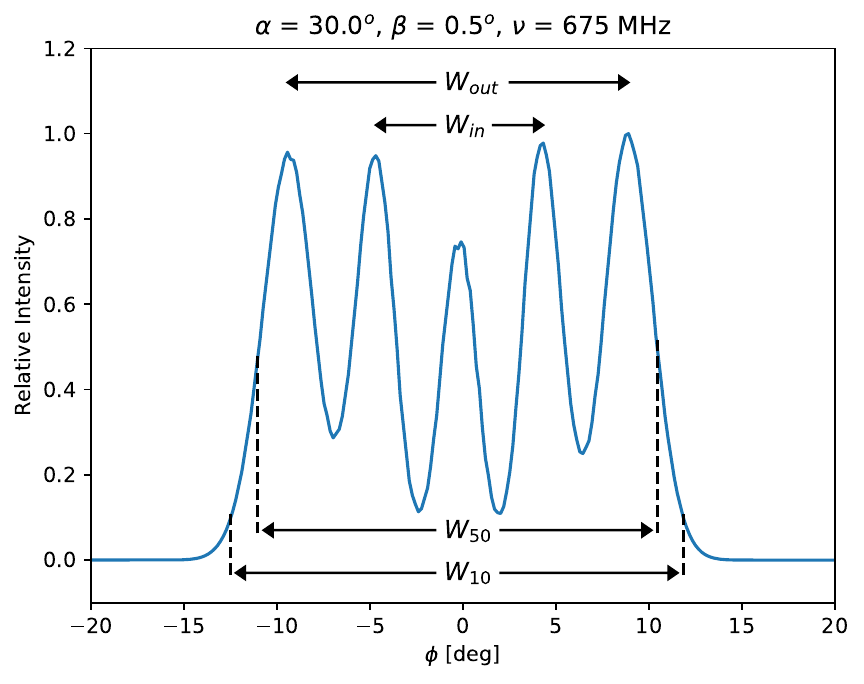}{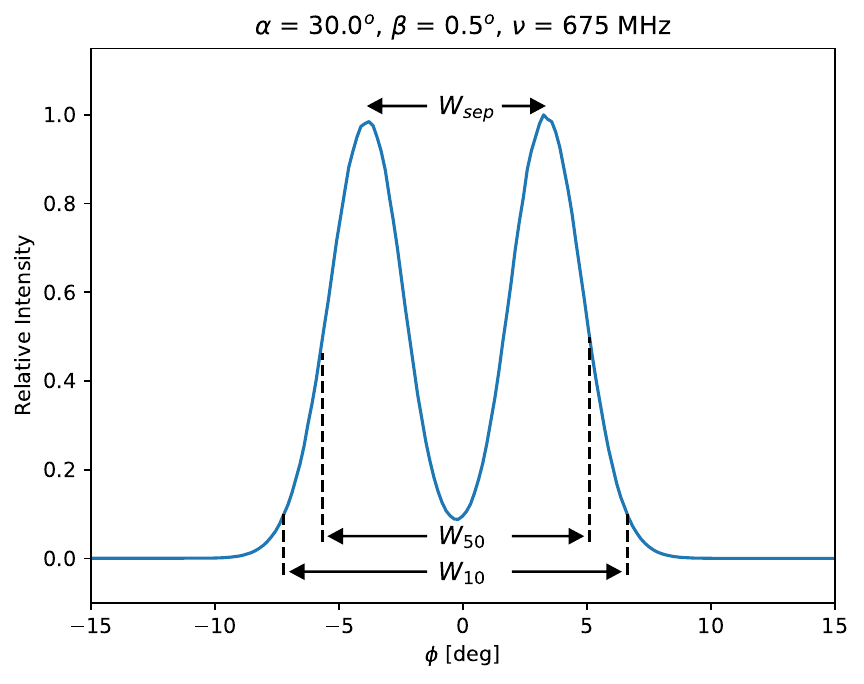}
\plottwo{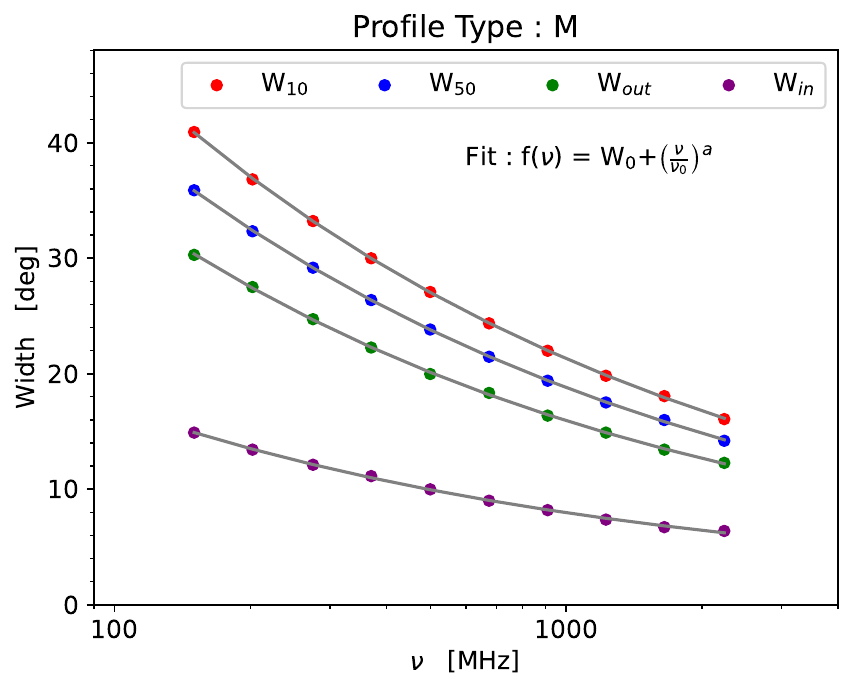}{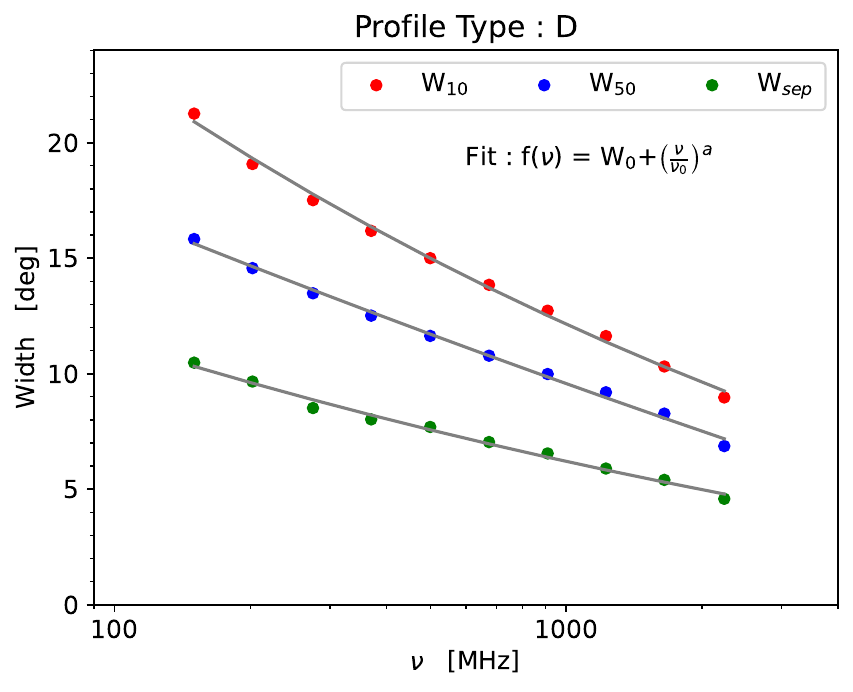}
\caption{The figure shows the evolution of estimated profile widths with 
frequency. The left panel corresponds to a M type profile and four different
widths have been measured, $W_{50}$, $W_{10}$, $W_{in}$ and $W_{out}$ (see top 
left panel). All widths decrease with increasing frequency, as shown in the 
bottom left panel, that has a functional form : $f(\nu) = W_0 + 
\left(\nu/\nu_0\right)^a$, where the coefficient $a$ is negative. The right 
panel shows the equivalent measurements for a D type profile where $W_{50}$, 
$W_{10}$ and $W_{sep}$ has been estimated as shown in the top right panel. The 
bottom right panel shows the frequency evolution of the widths in the D type 
profile.
\label{fig:RFM_wid}}
\end{figure}

Fig \ref{fig:RFM_ht} shows the estimated average profile (left window) and the 
grid intensity distribution (right window) at three frequencies, 150 MHz (top 
panel), 325 MHz (middle panel) and 1400 MHz (bottom panel). The pulsar radio
emission shows the effect of radius to frequency mapping \citep[RFM,][]{C78} 
where the emission at lower frequencies are expected to arise higher up the 
magnetosphere. The RFM is clearly seen in the radio emission from CCR due to 
charged solitons in the figure. The majority of the emission at 150 MHz arises 
from a range of heights between 400 km and 800 km. The range of heights 
decrease with increasing frequency, at 325 MHz the emission arises between 
200~km and 450 km, while at 1400 MHz the emission heights between 80~km and 
180~km. The core emission at the center emerges from a somewhat lower height 
compared to the cones, an effect that has also been predicted in certain 
analytical studies \citep[see Fig 5. in][]{MGM12}. At the lower frequencies the
shift is $\sim$ 50 -- 100~km and becomes smaller $\sim$ 20 -- 50~km at 1400 
MHz. The intensity of the core emission changes with respect to the cones as a 
function of frequency, where the core becomes comparatively less bright at 
higher frequencies (see section \ref{sec:spec} for more details).

\begin{deluxetable*}{cccccccc}
\tablecaption{Estimating the frequency evolution of profile widths 
\label{tab:profwid}}
\tabletypesize{\footnotesize}
\tablehead{
   & \multicolumn{4}{c}{Profile - M ($\beta = 0.5\degr$)} & \multicolumn{3}{c}{Profile - D ($\beta = 4.3\degr$)} \\
\tableline
 \colhead{Frequency} & \colhead{$W_{10}$} & \colhead{$W_{50}$} & \colhead{$W_{out}$} & \colhead{$W_{in}$} & \colhead{$W_{10}$} & \colhead{$W_{50}$} & \colhead{$W_{sep}$} \\
 \colhead{[MHz]} & \colhead{[deg]} & \colhead{[deg]} & \colhead{[deg]} & \colhead{[deg]} & \colhead{[deg]} & \colhead{[deg]} & \colhead{[deg]}}
\startdata
 150.0 & 40.9 & 35.9 & 30.3 & 14.9 & 21.3 & 15.8 & 10.5 \\
 202.0 & 36.8 & 32.3 & 27.5 & 13.4 & 19.1 & 14.6 &  9.7 \\
 275.0 & 33.2 & 29.2 & 24.7 & 12.1 & 17.5 & 13.5 &  8.5 \\
 370.0 & 30.0 & 26.4 & 22.3 & 11.1 & 16.2 & 12.5 &  8.0 \\
 500.0 & 27.1 & 23.8 & 20.0 & 10.0 & 15.0 & 11.6 &  7.7 \\
 675.0 & 24.4 & 21.5 & 18.3 &  9.0 & 13.8 & 10.8 &  7.0 \\
 910.0 & 22.0 & 19.4 & 16.4 &  8.2 & 12.7 & 10.0 &  6.5 \\
1225.0 & 19.8 & 17.5 & 14.9 &  7.4 & 11.6 &  9.2 &  5.9 \\
1650.0 & 18.1 & 16.0 & 13.4 &  6.7 & 10.3 &  8.3 &  5.4 \\
\tableline
   $a$ & $-0.34\pm0.01$ & $-0.34\pm0.01$ & $-0.36\pm0.02$ & $-0.38\pm0.03$ & $-0.27\pm0.07$ & $-0.19\pm0.04$ & $-0.29+\pm0.11$ \\
\enddata
\end{deluxetable*}

To further investigate the effect of RFM we have estimated the profiles of two
different systems at 9 frequencies between 150 MHz and 2 GHz (see Table 
\ref{tab:profwid}). The first case corresponds to the M profile discussed 
earlier, with two pairs of conal components around the central core. We have 
characterised the profile widths by four different methods, $W_{10}$ and 
$W_{50}$ representing the separation between the longitudes that are 10\% and 
50\% intensity levels of the outer most conal components, respectively. 
Additionally, the separation between the peak intensities of the inner and 
outer conal pairs, $W_{in}$ and $W_{out}$, are also estimated (see Fig. 
\ref{fig:RFM_wid}, top left panel). The second case is of a D type profile 
with $\beta = 4.3\degr$, $\alpha = 30\degr$, period of 1 seconds, and has one 
conal ring around the core component. The profile widths in the double peaked 
structures are measured at $W_{10}$ and $W_{50}$ levels from the two peaks as 
well as $W_{sep}$, measuring the longitudinal separation between them, as shown
in the top right panel of Fig. \ref{fig:RFM_wid}.

Fig \ref{fig:RFM_wid} also shows the variation of profile widths with frequency
for the M (bottom left window) and D (bottom right window) profiles, with clear
effects of RFM seen in their frequency evolution. \citet{T91} proposed a 
frequency dependence of profile widths : $W (\nu) = W_0 + A \nu^a$, where $W_0$
is the lower limit of width at high frequencies, $A$ is a proportionality 
constant and $a$ is the coefficient of frequency evolution. The last panel in 
Table \ref{tab:profwid} shows the estimates of $a$ obtained from the model fits
to different widths of the two systems. The frequency coefficient in most cases
is around -1/3, which is consistent with observations of several pulsars 
\citep{KG97,ET_MR02}.

\subsection {Geometrical Dependence of the Polarization Position Angle} \label{sec:PPA}

\begin{figure}
\gridline{\fig{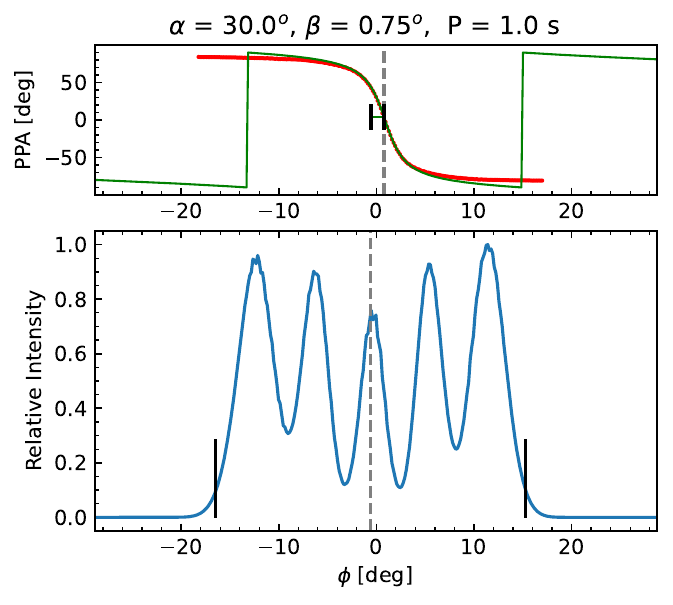}{0.48\textwidth}{}
         \fig{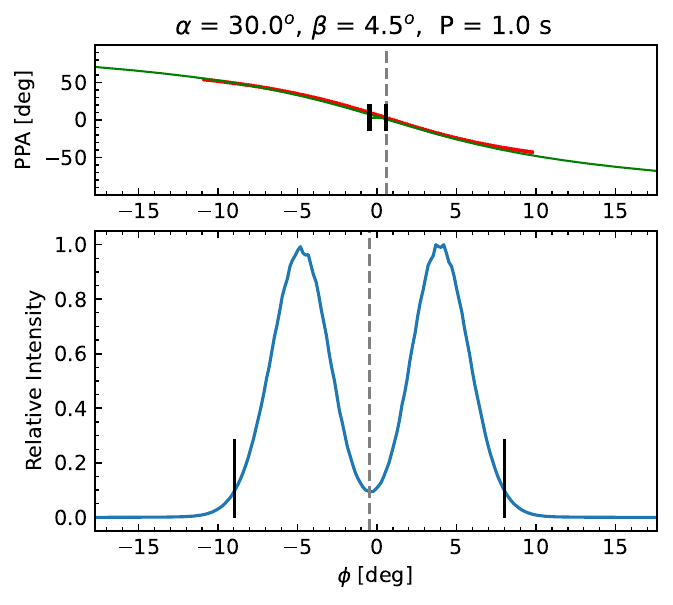}{0.48\textwidth}{}
         }
\gridline{\fig{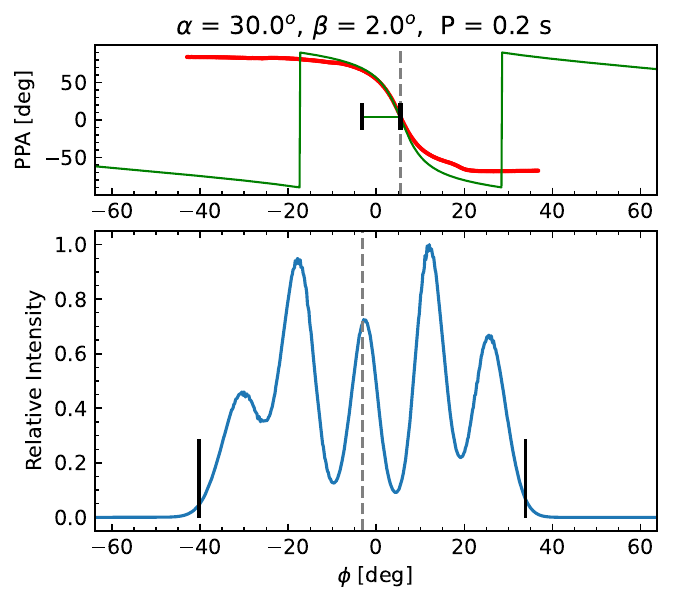}{0.48\textwidth}{}
         }
\caption{The PPA traverse across the profile is estimated at 325 MHz for three 
different configurations. The top left panel corresponds to a pulsar with 
rotation period of 1 second, inclination angle $\alpha=30\degr$ and a central 
line of sight cut, $\beta=0.75\degr$. The top right panel represents an outer 
line of sight cut for the same system with $\beta = 4.5\degr$. The lower panel 
shows the PPA variation in a faster rotating pulsar with period of 0.2 seconds,
$\alpha=30\degr$ and $\beta=2.0\degr$. The top window in each panel shows the 
estimated PPA (red line) from the average polarization obtained from the 
soliton grid, along with the expected PPA from the RVM (green line) using the 
respective geometrical angles. The RVM curve has been aligned with the measured
PPA in the longitudinal direction using the steepest gradient points (vertical 
dashed grey line in top window). The average profiles are shown in the bottom 
window along with the 10\% width levels on either side of the profiles (black 
vertical lines) and the center of the profile (vertical dashed grey line) as 
the mid point between the 10\% longitudes. The longitudinal separation, 
$\Delta\phi_H$, between the profile center and the steepest gradient point of 
PPA is shown as the horizontal green line in the center of the top windows.
\label{fig:RVMprof}}
\end{figure}

One of the distinguishing features of the pulsar radio emission is the PPA 
curve that exhibits a characteristic variation across the profile window. The
PPA ($\chi$) variation with LOS longitude ($\phi$) can be explained using RVM 
\citep{RC69}, where the characteristic curve can be derived from the emission 
geometry, specified by $\alpha$ and $\beta$ as :
\begin{equation}
\chi = \chi_o + \tan^{-1}{\left(\cfrac{\sin{\alpha} \sin{(\phi-\phi_o)}}
{\sin{(\alpha + \beta)} \cos{\alpha} - \sin{\alpha} \cos{(\alpha + \beta)}\cos{(\phi-\phi_o)}}\right)}.
\label{eq:RVM}
\end{equation}
Here, $\chi_o$ is the PPA value at the longitude, $\phi_o$, corresponding to 
the reference plane containing the rotation and magnetic axes. The reference 
longitude is expected to align with the steepest gradient (SG) point of the PPA 
curve where the slope is directly dependent on the geometrical angles :
\begin{equation}
\left(\cfrac{\partial\chi}{\partial\phi}\right)_{\phi=\phi_o} = \left(\cfrac{\sin{\alpha}}{\sin{\beta}}\right).
\label{eq:PPA_SG}
\end{equation}
The study of the PPA behaviour in pulsars is of particular importance as they
provide an independent measurement of the location of the radio emission in the 
magnetosphere. Due to the rotation motion of the star, the center of the 
profile is shifted to earlier longitudes (eq.\ref{eq:kvec}), while the fiducial 
plane gets delayed with respect to the center (eq.\ref{eq:PPA_ht}). The 
emission height ($h_E$) can be estimated using the separation ($\Delta\phi_H$) 
between the center of the profile and SG longitude of the PPA as \citep{BCW91,
ML04,D08} :
\begin{equation}
h_E = \frac{c P \Delta\phi_H}{8\pi} \approx ~208.2~\left(\frac{P}{\text{s}}\right) \left(\frac{\Delta\phi_H}{\text{deg}}\right)~\text{km}.
\label{eq:h_AR}
\end{equation}

\begin{deluxetable*}{ccccccccccccc}
\tablecaption{Estimating the Radio Emission Height from PPA Shift
\label{tab:PPA_ht}}
\tabletypesize{\small}
\tablehead{
 \colhead{Type} & \colhead{Frequency} & \colhead{$P$} & \colhead{$\alpha$} & \colhead{$\beta$} & \colhead{$\phi_l$} & \colhead{$\phi_t$} & \colhead{$\phi_c$} & \colhead{$\phi_o$} & \colhead{$|d\chi/d\phi|_{\phi_o}$} & \colhead{$\sin{\alpha}/\sin{\beta}$} & \colhead{$\Delta\phi_H$} & \colhead{$h_E$} \\
\tableline
   & \colhead{[MHz]} & \colhead{[s]} & \colhead{[deg]} & \colhead{[deg]} & \colhead{[deg]} & \colhead{[deg]} & \colhead{[deg]} & \colhead{[deg]} & \colhead{[deg/deg]} & \colhead{[deg/deg]} & \colhead{[deg]} & \colhead{[km]}}
\startdata
 M & 325 & 1.0 & 30.0 & 0.75 & -16.46 & 15.30 & -0.58 & 0.78 & 37.4 & 38.2 & 1.36 & 283 \\
   &  &  &  &  &  &  &  &  &  &  &  &  \\
 D & 325 & 1.0 & 30.0 & 4.5 & -8.94 & 7.99 & -0.47 & 0.59 & 6.76 & 6.37 & 1.06 & 221 \\
   &  &  &  &  &  &  &  &  &  &  &  &  \\
 M & 325 & 0.2 & 30.0 & 2.0 & -40.23 & 33.87 & -3.18 & 5.49 & 12.72 & 14.3 & 8.67 & 361 \\
\tableline
\enddata
\end{deluxetable*}

The RVM nature of the PPA tracks assumes the emission to arise from a narrow 
range of heights. However, in practical cases the emission at any given 
frequency is expected from a wider distance spanning several hundred kilometers 
(see section \ref{sec:RFM}). We investigate the effect of the averaging process
over a range of heights on the PPA nature of the emission. The average Stokes 
parameters, $U_L$ and $Q_L$, along every longitude, $\phi_L$, is calculated by 
adding the contribution of every soliton from eq. (\ref{eq:curv_spec2}) and 
(\ref{eq:curv_spec3}) along this direction, and the average PPA is estimated as
$\chi_L = 0.5\tan^{-1}(U_L/Q_L)$. Fig. \ref{fig:RVMprof} shows the pulsar 
profile and the average PPA variations for three different configurations at 
the frequency of 325 MHz and inclination angle $\alpha = 30\degr$. The top left
window represents a M type profile with a central LOS, $\beta = 0.75\degr$, and
rotation period of 1 second. In the top right window a more peripheral LOS with
$\beta = 4.5\degr$ is shown also with $P=1$ seconds, and forms a double 
component profile. The bottom window shows the emission features of a faster 
pulsar with $P=0.2$ seconds for a relatively central LOS, $\beta=2\degr$, 
forming a M type profile. The top window in each plot shows the estimated PPA 
(in red) from the emission along with the RVM fits (green line) obtained from 
eq. (\ref{eq:RVM}), with the relevant $\alpha$ and $\beta$ values, and $\phi_o$
obtained from the SG longitude of the PPA.

The SG longitude of the PPA, i.e. $\phi_o$, is estimated in each case and shown
as a vertical (dashed grey line) on the top window (also see Table 
\ref{tab:PPA_ht}). The average profile is shifted towards the leading side 
(negative longitudes), and the profile center can be identified as the mid 
point between a reference intensity level on either side. Table 
\ref{tab:PPA_ht} also lists $\phi_l$ located at 10\% peak intensity at the 
leading side of the first component, and $\phi_t$ is the equivalent longitude 
in the trailing side of the last component (see black vertical lines in each 
profile). This gives an estimate of the central longitude $\phi_c = (\phi_l + 
\phi_t)/2$, which is listed in the table for each pulsar configuration and also
shown in the profile plots as vertical dashed line. The shift between the 
profile center and the SG point, $\Delta\phi_H = \phi_o - \phi_c$, is larger in
the faster pulsar compared to the slower cases (see Table \ref{tab:PPA_ht}).

The PPA at any given emission height follows the RVM curve, however the RVM at 
different heights get shifted (see eq. \ref{eq:PPA_ht}), and after averaging 
across different heights the PPA behaviour deviates from the ideal RVM nature. 
The shift is more pronounced near the edges of the profile as averaging is over
a larger range of heights and covers more field lines due to their diverging 
nature. The deviation is also more pronounced in the shorter period pulsars 
since the cross section of the open field line region is wider. In principle 
the deviation between the measured PPA and the RVM should be visible in the 
pulsar observations. However, in practice the measurements are affected by 
telescope noise which are expected to be higher than the deviations in most 
cases \citep{MBM16,OMR19,MMB23b}. In addition, there are also orthogonal 
polarization modes, where during certain instances either the parallel or 
perpendicular component of the intensity gets absorbed during propagation in 
the pulsar plasma \citep{MMB23a,MBM24b}. This causes further deviations of the 
observed average PPA from RVM.

The SG point of the PPA is less affected by the averaging and appears to match 
the ideal RVM behaviour, with the slopes having 10\% difference for the faster 
pulsar and less than 5\% difference for the two longer period pulsars. The 
emission height estimates varies between 220 km for the outer LOS cut, to 280 
km for the slower pulsar with M profile and 360 km in case of the faster 
pulsar. All estimates are consistent with the range of heights, between 200 -- 
450~km, where the solitons emitting CCR at 325 MHz are located within the open 
field line region (see Fig \ref{fig:RFM_ht}, middle panel). These results 
highlight that at any given frequency, despite the averaging over a range of 
emission heights, the RVM fits to the average PPA, particularly the estimation 
of the SG point, is a useful tool to study the physical properties of the 
pulsar.

\subsection{Spectral Nature of Pulsar Radio Emission and Variation Across the 
Profile} \label{sec:spec}

\begin{figure}
\gridline{\fig{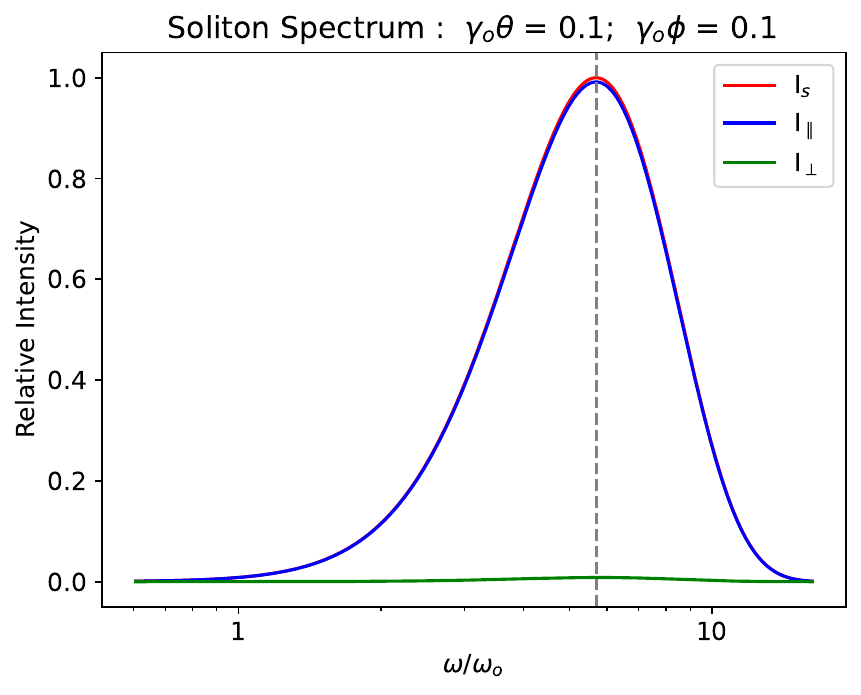}{0.48\textwidth}{}
          \fig{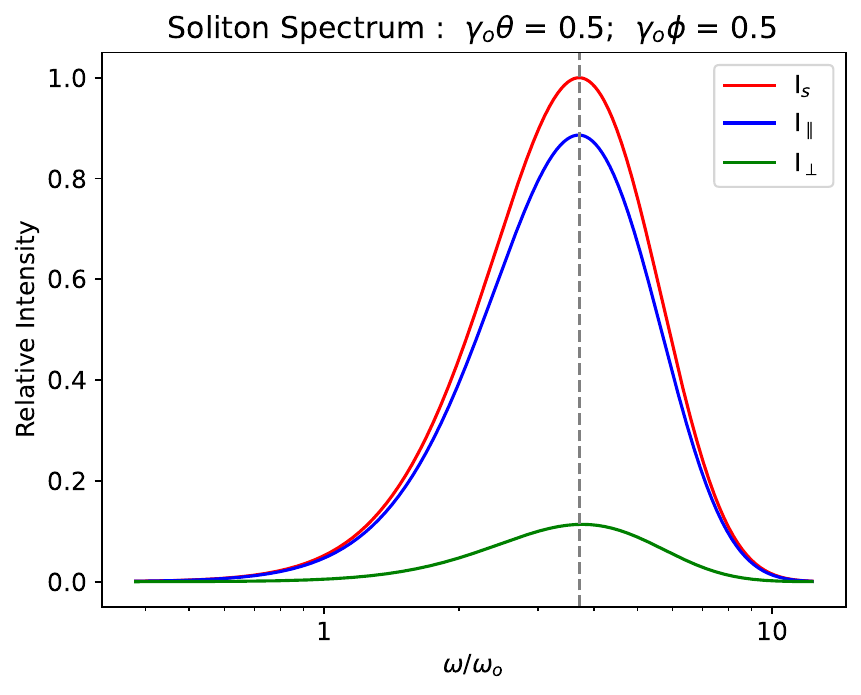}{0.48\textwidth}{}
         }
\gridline{\fig{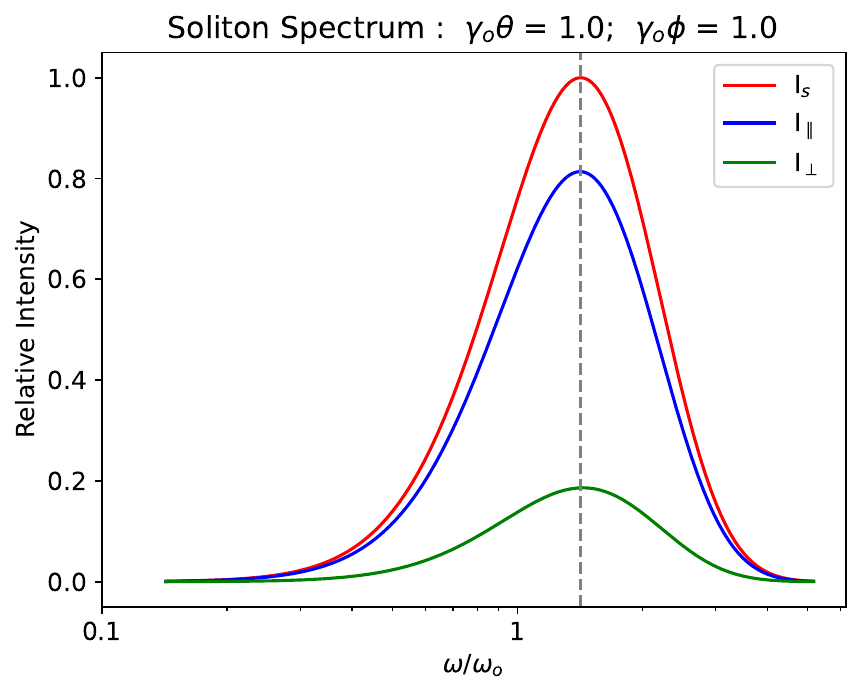}{0.48\textwidth}{}
         }
\caption{The spectrum of curvature radiation from soliton at different angular 
cuts across the emission beam ($\sim1/\gamma_o$), with the angular co-ordinates 
normalized by the beaming angle.
\label{fig:sol_spec}}
\end{figure}

The pulsar emission is distinguishable due to its spectral behaviour, which 
usually resemble an inverted power law function, $S\propto\nu^{b}$, with a 
steep spectral index, typically around $b\sim-1.6$ \citep{MKK00,JvK18}. The 
power law spectral nature is usually seen over a relatively wide frequency 
range between 100 MHz and up to few GHz. The spectra of many pulsars show a low
frequency turnover around 100 MHz \citep{IKM81}, and in around 40 cases the 
gigahertz peaked spectrum (GPS) behaviour is seen with a turnover between 
500~MHz and 1~GHz \citep{KLM11,KBL21}. The GPS nature is primarily attributed 
to the pulsar emission being absorbed in the intervening medium, while the 
remaining features should follow from the emission mechanism.

The spectral nature of curvature radiation from an individual soliton, shown in
Fig \ref{fig:sol_spec}, is a peaked structure at a certain frequency, whose 
intensity level drops off rapidly on either side of the peak. The location of 
the peak intensity depends on the angular distance from the tangent along the 
propagation direction. The emission is confined within a narrow angular cone 
defined by the Lorentz factor of solitons, $\delta\theta \leq 1.5/\gamma_o$ 
(see Fig \ref{fig:sol_prop}). Fig \ref{fig:sol_spec} shows the characteristic 
spectrum of curvature radiation from solitons using typical value of $a_s = 
0.3$ (see eq.\ref{eq:sol_spec}) at three different angular distances specified 
by $\gamma_o\theta = \gamma_o\phi=$ 0.1, 0.5 and 1.0. The spectrum represents 
the total intensity as well as the parallel and perpendicular field 
intensities. Near the tangential axis (top left window) the peak frequency is 
around $\omega/\omega_o\sim$ 5 -- 6 and is dominated by the parallel component. 
With increasing angular separation the peak intensity shifts towards $\omega_o$
and has larger contribution from the perpendicular component.

\begin{figure}
\gridline{\fig{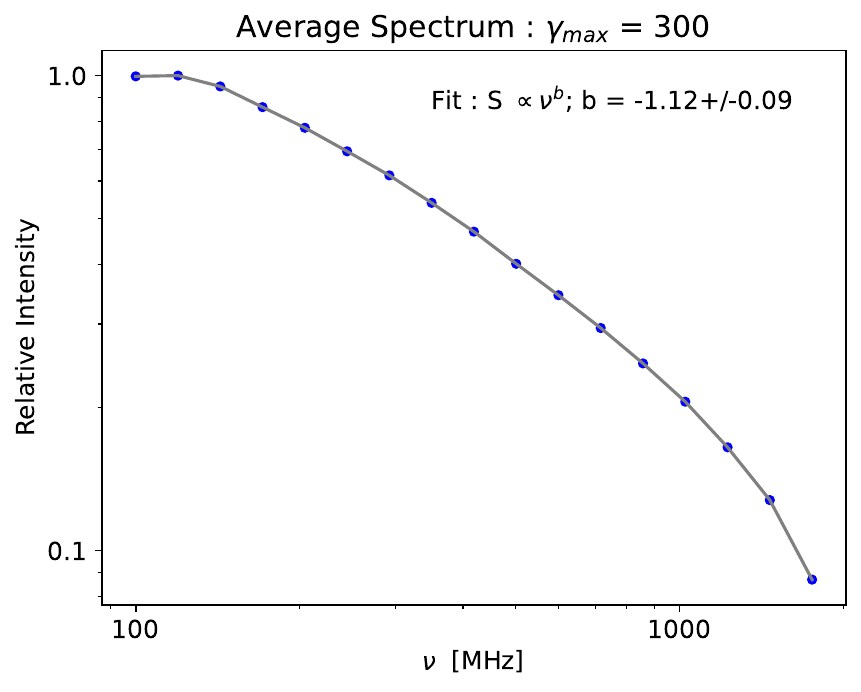}{0.48\textwidth}{}
          \fig{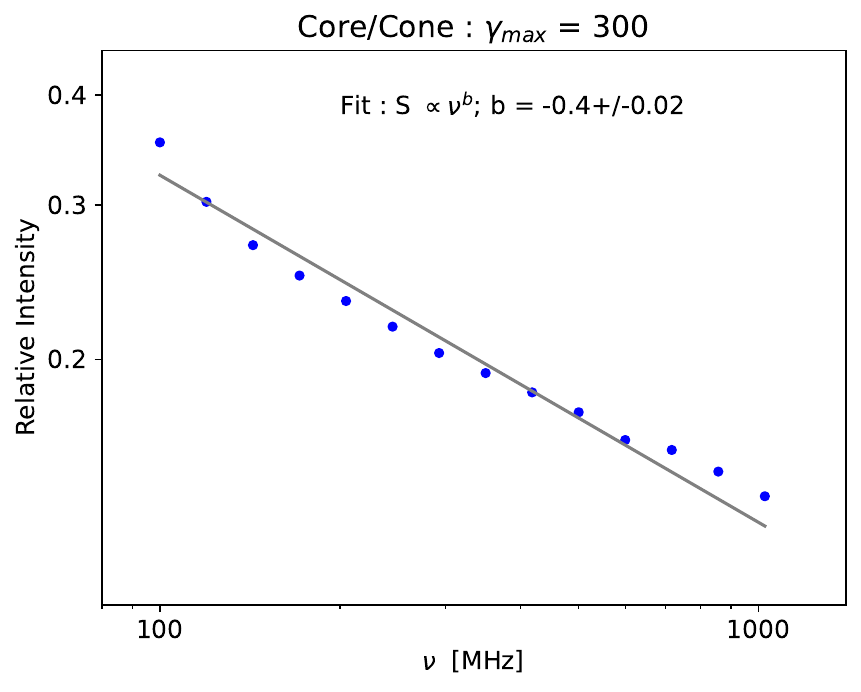}{0.48\textwidth}{}
         }
\gridline{\fig{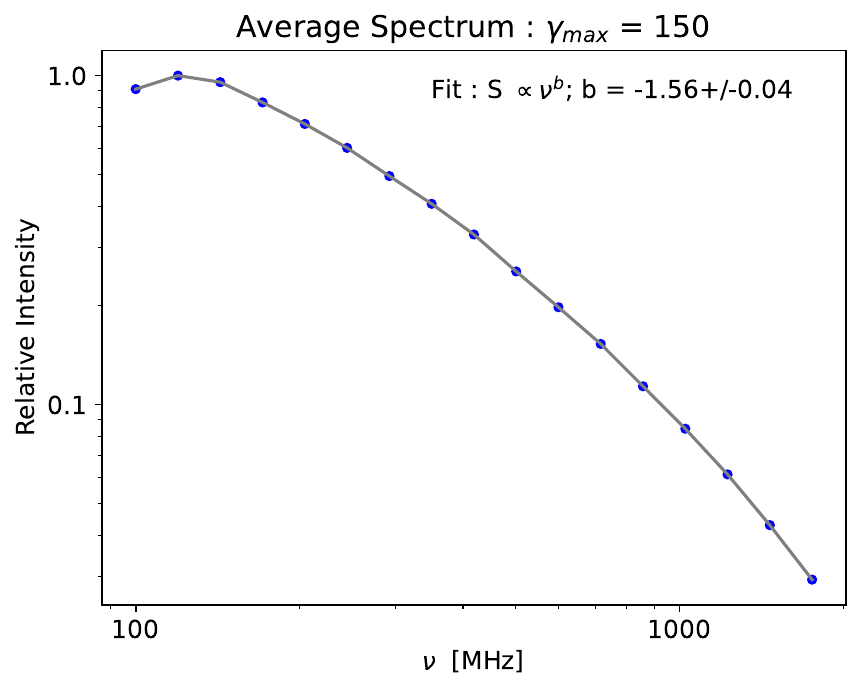}{0.48\textwidth}{}
          \fig{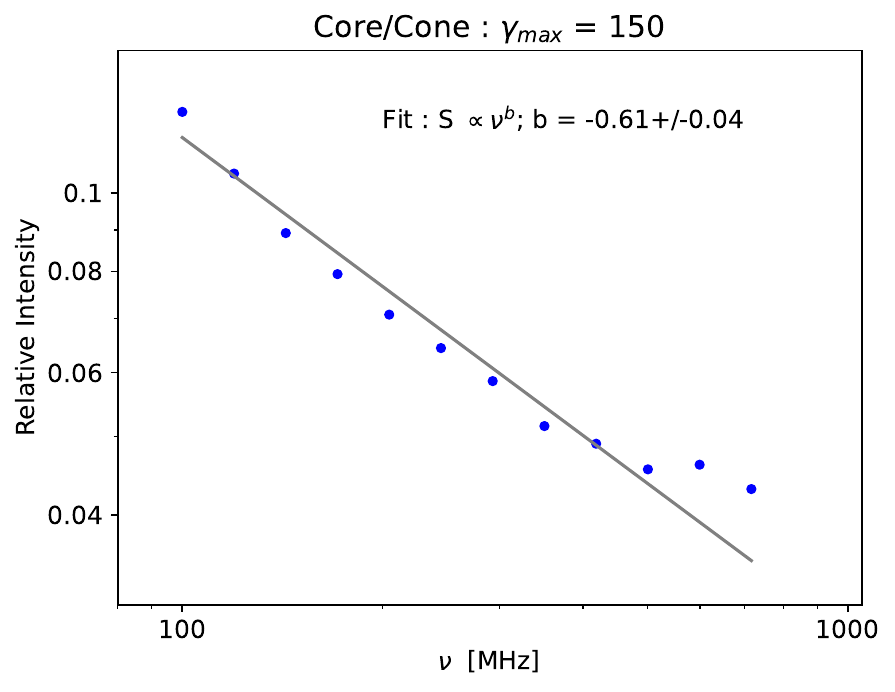}{0.5\textwidth}{}
         }
\caption{The estimated pulsar spectrum is shown in the left panel for two 
different secondary plasma distributions, in the frequency range between 100 
MHz and 1400 MHz. The right panel shows the ratio of the core and conal 
intensities as a function of frequency.
\label{fig:spec_puls}}
\end{figure}

The average emission of the pulsar is obtained by adding contributions from 
large number of solitons in the open field line region along the LOS. Fig 
\ref{fig:RFM_ht} shows that the emission at lower frequencies originates from
a wider range of emission heights compared to higher frequencies. This ensures 
that more solitons contribute at lower frequencies forming the inverted power 
law spectrum. Observations report the total intensity spectra of pulsars, which
show an inverted power law behaviour between 100 MHz and few GHz. To compare 
the spectral nature of CCR from charged solitons with observed behaviour, 
pulsar profiles are estimated as regular intervals between 100 MHz and 2 GHz. 
We consider a central LOS traverse with $\beta =0.5\degr$, inclination angle 
$\alpha = 30\degr$ in a pulsar with period of 1 second, resulting in a M type 
profile with core and conal emission. Fig \ref{fig:spec_puls} shows the 
spectral nature of the average curvature radiation for two different secondary 
plasma distributions, $\gamma_s$ between 50 and 300 (top panel) and $\gamma_s$ 
between 50 and 150 (bottom panel). The inverted power law spectrum is seen with
spectral index of $b\sim-1.1$ in the first case and $b\sim-1.6$ for the second 
distribution, which matches the observed behaviour in pulsars. In addition, the
low frequency turnover around 100 MHz is also seen in both spectra, which 
arises due to the upper limits in the energy distribution of the secondary 
plasma. 

Observational studies have further shown that the components within the pulsar
profile have different frequency evolution \citep{ET_R83,BMM21}. The core 
component has a steeper spectrum than the cones and is seen more prominently at
lower frequencies, while the conal components appear comparatively brighter at
the high frequency range. \citet{BMM22a} measured the relative spectral indices
between the core and the conal emission in a large sample of T and M type 
profiles and found the core spectra to have a steeper spectral index that 
cones. The differences in the spectral indices, $\Delta b$, have a nearly 
uniform distribution between $-0.2$ and $-2.0$, with slight peak around $\Delta
b \sim -0.7$. It was further shown that the CCR from charged solitons is pulsar 
plasma is able to explain the observed difference between the core and conal 
spectra. Although, the numerical estimates in this work was rather rudimentary.
A two dimensional configuration was considered for the pulsar magnetosphere and
the effect of pulsar rotation was not included while determining the direction 
of emission. 

The more detailed calculations in this work also reveal the core emission to
become less prominent a higher frequencies, as seen in the evolution of the
profile shape with frequency in Fig \ref{fig:RFM_ht}, left windows. This 
results from two contributing factors; firstly, as the emission heights 
decrease with increasing frequencies the LOS cut for the same $\beta$ is 
further away from the center, such that the intensity decreases \citep{S97}. 
Secondly, $\mathcal{R}$ increases with height (see right window of Fig 
\ref{fig:rad_curv} in Appendix \ref{app:radcurv}), such that to obtain the same
characteristic frequency, $\omega_o \sim \gamma_o^3/\mathcal{R}$, the Lorentz 
factors need to be higher. The intensity of emission is proportional to 
$\gamma_o^2$ (see eq. \ref{eq:curv_spec1} -- \ref{eq:curv_spec4}) and increases 
with decreasing frequency. We estimated the frequency evolution of the ratio 
between the core and conal intensities in the two M type profiles discussed 
earlier for the average spectral studies (see Fig \ref{fig:spec_puls}). The 
relative difference in the spectral indices are -0.4 and -0.6, respectively, 
that once again conform with observational expectations. The core emission 
becomes weak and mostly non-detectable above frequencies of 1~GHz, thereby 
showing the deviations from the linear nature at higher frequencies.

\section{Discussion} \label{sec:dis}

Several decades of observations have established that the radio emission from
the normal pulsar population, with certain exceptions, exhibit specific 
tendencies. These include the nature of the average profile and the concept of
an emission beam, the evolution of the profile and component widths with 
different parameters, the location of the radio emission region within the 
magnetosphere, the polarization and PPA behaviour, the spectral nature, amongst
several others. These features are symptomatic of the radio emission mechanism
in pulsars and are expected to emerge from similar physical conditions. 
Although there are ample studies addressing the origin of specific 
observational features, there is a dearth of detailed works connecting 
different aspects of the emission from similar physical framework. 

Detailed observations have demonstrated that radio emission arises due to a 
coherent emission mechanism between 100 -- 1000 kilometers from the stellar 
surface. The CCR from charge separated solitons, formed due to non-linear 
plasma instabilities, is the only known coherent mechanism that can develop at 
these altitudes. The primary objective of this work is to estimate different 
emission features that arise from a steady distribution of solitons in the 
pulsar plasma and find comparisons with the observations. We have studied four 
distinct emission behaviour that emerge from this mechanism :
\begin{enumerate}[(a)]
\item The nature of the average profiles and their relationship with LOS 
geometry,
\item The location of radio emission within the open field line region and 
their variation with emission frequency,
\item The nature of the emergent PPA in the profile after averaging over the 
range of emission altitudes, 
\item The spectral nature of the emission intensity and their variations across
different profile components.
\end{enumerate}
The solitons in the secondary plasma generated by a distribution of sparks in
the IAR reproduces the different profile types seen, where the nature of the 
profile depends on the LOS traverse, i.e. value of the angle $\beta$. The 
effect of radius to frequency mapping seen in pulsar observations (Maciesiak et
al. in preparation), where the lower frequency emission arises from higher 
altitudes, are clearly evident in the profile estimates. We also find the PPA 
variations to be largely consistent with the RVM nature, with the steepest 
gradient point being fairly well described by RVM, while there is some 
deviation near the conal wings of the PPA, particularly in faster pulsars. The 
average spectra from CCR due to charged solitons also show the spectral index 
to be similar to the median value of the pulsar population and clearly 
reproduce the steep spectral nature of the cores compared to the conal 
components. In all these estimates we used identical range of physical 
parameters, like the energy distribution of secondary plasma, the range of 
soliton sizes, etc. These results clearly demonstrate that CCR from charged 
solitons is able to produce the different observational features and is a 
strong candidate for the radio emission mechanism in pulsars.

In this work we have primarily focussed on the average emission properties that
can be explained using a steady state distribution of solitons along the open
filed line region, and are primarily determined by the LOS geometry and the
range of altitudes where the favourable conditions for the relevant plasma 
instabilities can develop. Other features can also be explained using the same 
setup when specific effects are considered, like the presence of non-dipolar 
surface fields, temporal variations introduced by the non-stationary sparking 
process as well as subpulse drifting, the plasma effects due to the density 
variations in the secondary plasma, both in the longitudinal and transverse 
directions, the effect of positively charged ions on the outflowing plasma, 
etc. For example the non-dipolar fields can explain the asymmetric emission 
seen across the emission window of several pulsars. The presence of 
microstructures in the single pulses can be explained using the fluctuations of
the secondary plasma density in the longitudinal direction \citep{MBM20}. 
The detailed polarization behaviour like presence of orthogonal polarization 
modes, time samples with high levels of linear polarization, the circular 
polarization, requires introducing propagation effects in inhomogeneous plasma. 

In this work we have assumed vacuum like propagation of the emission modes 
excited in the plasma which has certain limitations. In the pulsar plasma it is
expected that the CCR from charged solitons excites the transverse, $t$-mode 
and longitudinal-transverse, $lt_1$-mode in the strongly magnetized pair 
plasma, and due to different refractive indices the modes splits and propagate 
independently \citep{AB86}. The $t$-mode has vacuum like properties and can 
escape the plasma and emerge as the extraordinary (X) waves. The $lt_1$-mode is
ducted along field lines and usually gets damped, but under certain conditions 
can emerge as the ordinary (O) waves. Thus several possibilities exist for the 
emergent radiation from the plasma before it reaches the observer. For example 
if the $lt_1$-mode is entirely damped, then the radiation is only composed of 
the X-mode, and show high levels of linear polarization. In this case most of 
the properties such as RVM, spectral variation, PPA shifts, etc., presented in 
this work would remain unaffected, with the polarization vector directed 
perpendicular to the magnetic field line planes. Another possibility is that 
the $lt_1$-mode is only partially damped, and hence an incoherent mixing of 
X-mode and O-mode can escape the plasma, where the O-mode may also be affected 
due to refraction in the plasma boundary \citep[see][for a more detailed 
discussion]{MBM24b}. Understanding each of these effects within the framework 
of solitons specified in this work would require dedicated studies and will be 
addressed in future works.

\section*{Acknowledgments}
We thank the anonymous referee for comments that helped to improve the paper. 
DM acknowledges the support of the Department of Atomic Energy, Government of
India, under project no. 12-R\&D-TFR-5.02-0700. This work was supported by the
grant 2020/37/B/ST9/02215 of the National Science Centre, Poland.

\bibliography{reflist}{}
\bibliographystyle{aasjournal}

\appendix
\section{Radius of Curvature of Dipolar Magnetic field}\label{app:radcurv}

The general formalism for the radius of curvature ($\mathcal{R}$) of the 
magnetic field lines from a system of dipoles has been shown in earlier works 
\citep{GMM02,BMM20b}. In this section we show the estimates of $\mathcal{R}$ 
for purely dipolar fields in pulsar emission region and their variation with
emission altitude as well as in the azimuthal direction across the open field 
line region. The general form of the dipolar magnetic field in spherical 
co-ordinates, $\bm{B} = (B_r, B_\theta, B_\phi)$ is represented in 
eq.(\ref{eq:magdip}), with the dipolar axis inclined by an angle $\alpha$ with
the $z$-axis, which is along the rotation axis. The line of force can be 
estimated from the magnetic field as :
\begin{equation}
\frac{\mathrm{d}\theta}{\mathrm{d}r} = \cfrac{B_\theta}{r B_r} ~ \equiv ~ \Theta_1, ~~~~ \frac{\mathrm{d}\phi}{\mathrm{d}r} = \cfrac{B_\phi}{r B_r \sin{\theta}} ~ \equiv ~\Phi_1.
\label{eq:fieldline}
\end{equation}
The curvature ($\mathcal{K} = 1/\mathcal{R}$) of the magnetic field line has 
the form :
\begin{equation}\label{eq:curv1}
\mathcal{K} = \left(\frac{\mathrm{d}s}{\mathrm{d}r}\right)^{-3}  \left\vert \left(\frac{\mathrm{d}^2\bm{r}}{\mathrm{d}r^2}\frac{\mathrm{d}s}{\mathrm{d}r} - \frac{\mathrm{d}\bm{r}}{\mathrm{d}r}\frac{\mathrm{d}^2 s}{\mathrm{d}r^2}\right) \right\vert,
\end{equation}
such that $\mathcal{R} = S_1^3 (J_1^2 + J_2^2 + J_3^2)^{-1/2}$, where $J_1 = 
X_2 S_1 - X_1 S_2$; $J_2 = Y_2 S_1 - Y_1 S_2$ and $J_3 = Z_2 S_1 - Z_1 S_2$, 
with $S_1 = \mathrm{d}s/\mathrm{d}r$ and $S_2 = \mathrm{d}^2s/\mathrm{d}r^2$. 
The individual terms can be expressed in terms of the different co-ordinates as :
\begin{equation}\label{eq:curv2}
\begin{split}
X_1 & = \sin{\theta}\cos{\phi} + r\Theta_1\cos{\theta}\cos{\phi} - r\Phi_1\sin{\theta}\sin{\phi},\\
Y_1 & = \sin{\theta}\sin{\phi} + r\Theta_1\cos{\theta}\sin{\phi} + r\Phi_1\sin{\theta}\cos{\phi}, \\
Z_1 & = \cos{\theta} - r\Theta_1\sin{\theta}, \\
X_2 & = (2\Theta_1 + r\Theta_2)\cos{\theta}\cos{\phi} - (2\Phi_1 + r\Phi_2)\sin{\theta}\sin{\phi} - r (\Theta_1^2 + \Phi_1^2)\sin{\theta}\cos{\phi} - 2r\Theta_1\Phi_1\cos{\theta}\sin{\phi}, \\
Y_2 & = (2\Theta_1 + r\Theta_2)\cos{\theta}\sin{\phi} + (2\Phi_1 + r\Phi_2)\sin{\theta}\cos{\phi} - r(\Theta_1^2 + \Phi_1^2)\sin{\theta}\sin{\phi} + 2r\Theta_1\Phi_1\cos{\theta}\cos{\phi}, \\
Z_2 & = -2\Theta_1\sin{\theta} - r\Theta_2\sin{\theta}-r\Theta_1^2\cos{\theta}, \\
S_1 & = (1 + r^2\Theta_1^2 + r^2\Phi_1^2\sin^2{\theta})^{1/2}, \\
S_2 & = S_1^{-1} (r\Theta_1^2 + r^2\Theta_1\Theta_2 + r\Phi_1^2\sin^2{\theta} + r^2\Phi_1\Phi_2\sin^2{\theta} + r^2\Theta_1\Phi_1^2\sin{\theta}\cos{\theta}). 
\end{split}
\end{equation}
In dipolar fields the second order derivatives $\Theta_2 = 
\mathrm{d}\Theta_1/\mathrm{d}r$ and $\Phi_2 = \mathrm{d}\Phi_1/\mathrm{d}r$ has
the form :
\begin{equation}\label{eq:curv3}
\begin{split}
\Theta_2 = & - \frac{\Theta_1}{r} + \frac{1}{r B_r} \frac{\mathrm{d}B_\theta}{\mathrm{d}r} - \frac{\Theta_1}{B_r} \frac{\mathrm{d}B_r}{\mathrm{d}r} \\
\Phi_2 = & - \frac{\Phi_1}{r} + \frac{1}{r B_r \sin{\theta}} \frac{\mathrm{d}B_\phi}{\mathrm{d}r} - \frac{\Phi_1}{B_r}\frac{\mathrm{d}B_r}{\mathrm{d}r} - \Theta_1 \Phi_1 \cot{\theta},
\end{split}
\end{equation}
and the derivatives of the magnetic field components can be written as :
\begin{equation}\label{eq:curv4}
\begin{split}
\frac{\mathrm{d}B_r}{\mathrm{d}r} & = - \frac{3B_r}{r} - 2B_\theta\Theta_1 - 2B_\phi\sin{\theta}~\Phi_1, \\
\frac{\mathrm{d}B_\theta}{\mathrm{d}r} & = ~~~\frac{3B_\theta}{r} + \frac{1}{2}B_r\Theta_1 + B_\phi\cos{\theta}~\Phi_1, \\
\frac{\mathrm{d}B_\phi}{\mathrm{d}r} & = - \frac{3B_\phi}{r} + B_\phi\cot{\phi}~\Phi_1. \\
\end{split}
\end{equation}

\begin{figure}
\epsscale{1.15}
\plottwo{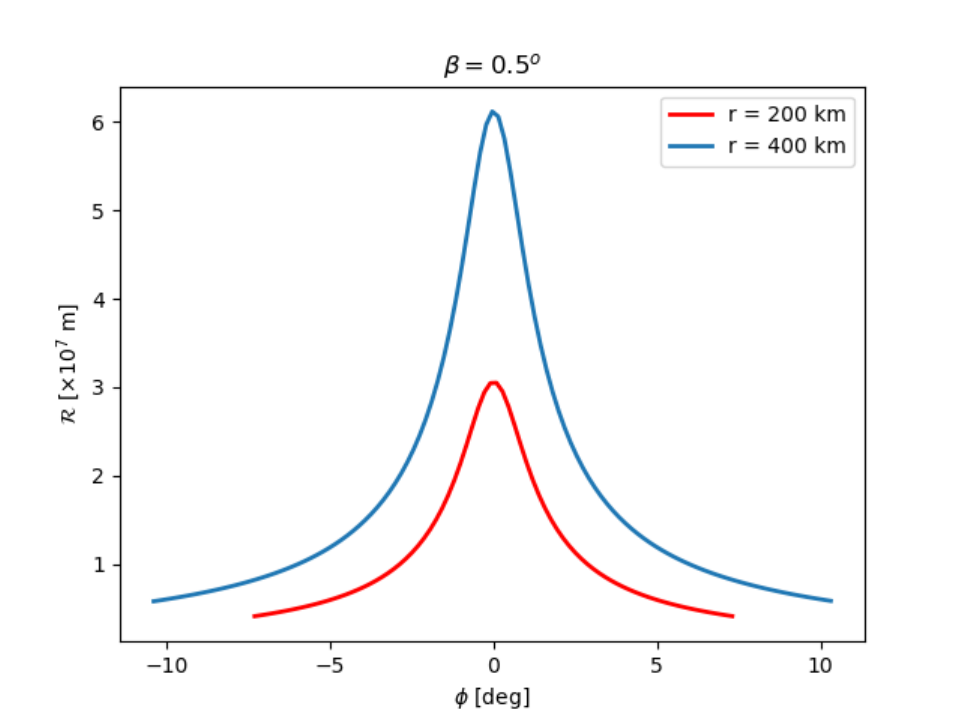}{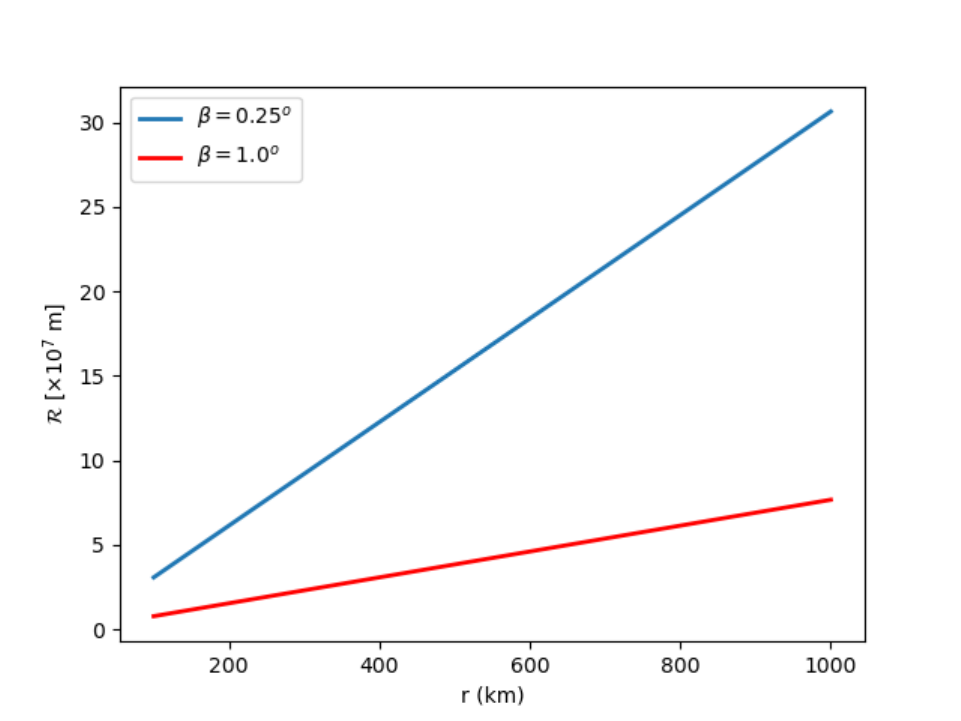}
\caption{The figure shows the radius of curvature ($\mathcal{R}$) of the 
dipolar magnetic field lines as a function of the co-ordinates. (a) The 
variation across the open field line region along a specific line of sight, 
$\beta = 0.5\degr$, at heights of $r=200$ km (red) and $r=400$ km (blue). (b)
The variation of between height range of $r=100$ km and $r=1000$ km, at two 
specific line of sights specified by $\beta = 0.25\degr$ (blue) and $\beta = 
1.0\degr$ (red).
\label{fig:rad_curv}}
\end{figure}

In Fig.\ref{fig:rad_curv} the variations of $\mathcal{R}$ across the open field
line region and as a function of distance from the star center is shown using
the estimates provided above.

\section{Two-dimensional curvature radiation spectrum} \label{app:curv_spec}

\begin{figure}
\epsscale{0.55}
\plotone{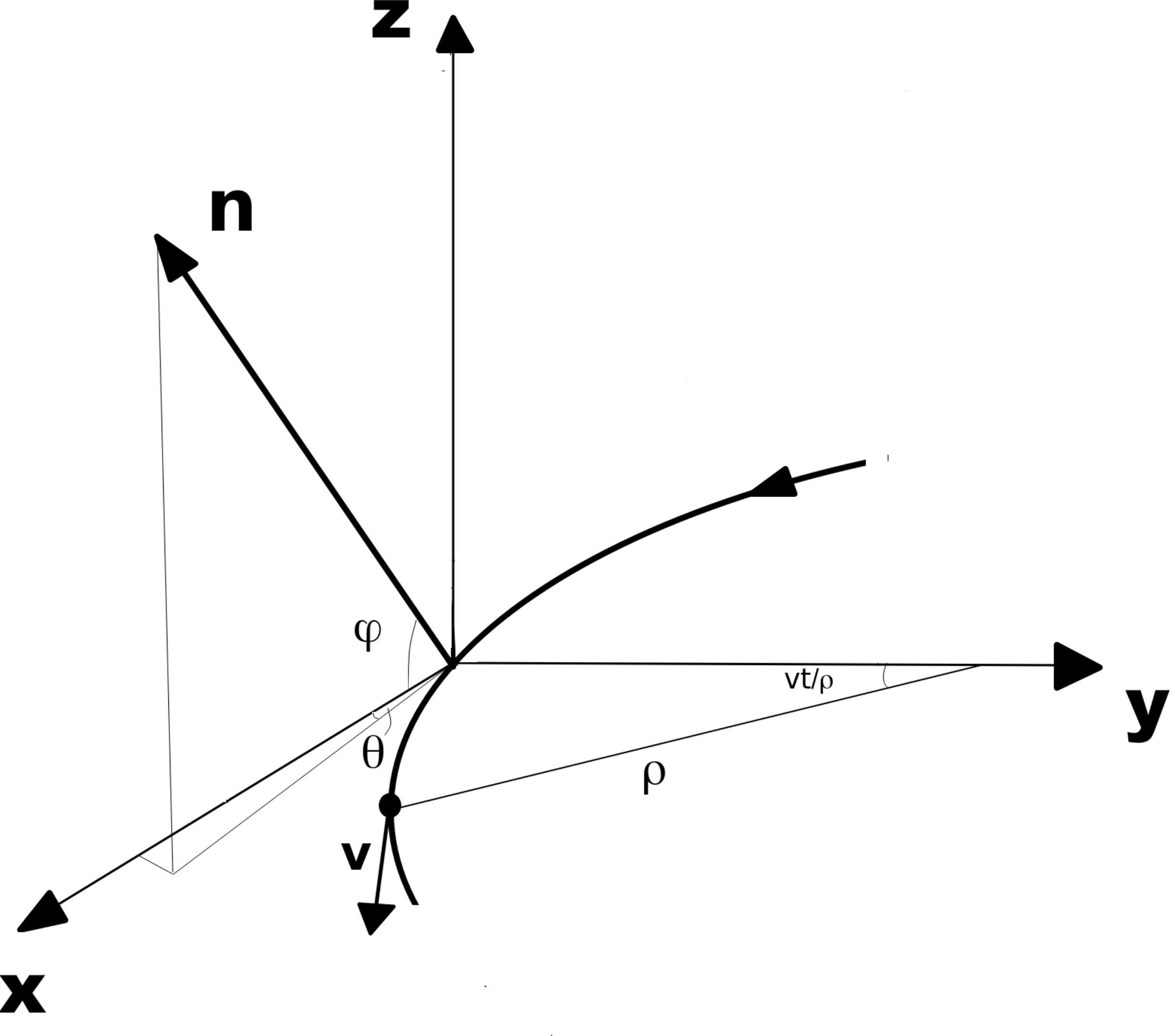}
\caption{The schematic shows co-ordinate setup for estimating the curvature
radiation from a charged particle moving along the open field line in the 
pulsar frame. The curved trajectory is considered to be in the $x-y$ plane with
a radius of curvature $\mathcal{R}$ and velocity vector of the particle at any 
instant specified by $\bm{v}$. The observer direction is along $\bm{n}$ 
specified by the angle $\theta$ with the $x$-axis in the $x-y$ plane and angle 
$\pi/2 - \phi$ with the $z$-axis.
\label{fig:curv_cord}}
\end{figure}

The energy ($I$) emitted by a relativistic particle moving along a curved 
trajectory, per unit solid angle ($S$) and per unit frequency ($\omega$) has 
the form \citep[see][]{J98}
\begin{equation}
\frac{dI}{d\omega d S}=\frac{e^2\omega^2}{4\pi^2c}\left|\int_{-\infty}^{+\infty}\bm{n}\times(\bm{n}\times\bm{\beta_{\rm curv}})e^{i\omega(t-\bm{n}\cdot\bm{r}(t)/c)}dt\right|^2.
\label{eqn:dIdf}
\end{equation}
Here, $\bm{r}(t)$ is the position of the particle at any instant $t$, 
$\bm{\beta_{\rm curv}} = \bm{v}/c$, corresponds to the velocity of the particle
and $\bm{n}$ is the unit vector directed towards the observer. In most 
textbooks $dI/d\omega d S$ variations are usually found along a single 
angular co-ordinate. The numerical estimation of the pulsar emission features
requires a more general form along both angular co-ordinates, which we evaluate
in this section \citep[also see appendix in ][]{YZ18}. Fig.\ref{fig:curv_cord}
presents the choice of the co-ordinate system, which is centered at the 
instantaneous position of the particle, and the particle trajectory is located
in the $x-y$ plane. The radius of curvature of the magnetic field line is 
$\mathcal{R}$ and at a any given instant $t$ the velocity vector makes an angle 
$v t/\mathcal{R}$ with the $y$-axis, such that :
\begin{eqnarray}
\bm{r}(t) = \mathcal{R} \sin{\left(\frac{v t}{\mathcal{R}}\right)}~\bm{\hat{x}} - \mathcal{R} \cos{\left(\frac{v t}{\mathcal{R}}\right)}~\bm{\hat{y}} \nonumber \\
\bm{\hat{v}} = \cos{\left(\frac{v t}{\mathcal{R}}\right)}~\bm{\hat{x}} + \sin{\left(\frac{v t}{\mathcal{R}}\right)}~\bm{\hat{y}}
\end{eqnarray}
The observers line of sight along $\bm{n}$ specified by the angles $\theta$ and
$\phi$ has the form :
\begin{equation}
\bm{n} = \cos{\theta}\cos{\phi}~\bm{\hat{x}} + \sin{\theta}\cos{\phi}~\bm{\hat{y}} + \sin{\phi}~\bm{\hat{z}}
\end{equation}
The emission intensity in equation (\ref{eqn:dIdf}) can be expressed in terms 
of two perpendicular amplitudes 
\begin{equation}
\frac{dI}{d\omega d S}=\left|\mathcal{E}_\parallel\bm{\epsilon}_\parallel + \mathcal{E}_\perp\bm{\epsilon}_\perp\right|^2
\end{equation}
where $\bm{\epsilon}_\parallel = \bm{\hat{y}}$,	is in the plane of the curved 
trajectory along the direction of acceleration, and $\bm{\epsilon}_\perp = 
\bm{n}\times\bm{\epsilon}_\parallel$. The radiation is beamed in a narrow cone
and hence the two amplitudes $\mathcal{E}_\parallel$ and $\mathcal{E}_\perp$ 
can be obtained using small angle approximation for $\theta$ and $\phi$. We 
also consider the particle to be ultra-relativistic hence $\beta_{\rm curv} 
\sim 1$. Using these approximations the exponential term in eq.(\ref{eqn:dIdf})
is represented as :
\begin{equation}
\omega\left(t-\frac{\bm{n}\cdot\bm{r}(t)}{c}\right) \simeq \frac{\omega}{2}\left[\left(\frac{1}{\gamma^2}+\theta^2+\phi^2\right)t+\frac{c^2t^3}{3\mathcal{R}^2}-\frac{ct^2}{\mathcal{R}}\theta\right],
\end{equation}
and the two perpendicular amplitudes take the form :
\begin{eqnarray}
\mathcal{E}_{\parallel} & \simeq & \mathcal{E}_o \omega \int_{-\infty}^{\infty}\left(\frac{ct}{\mathcal{R}}-\theta\right)\exp\left(i\frac{\omega}{2}\left[\left(\frac{1}{\gamma^2}+\theta^2+\phi^2\right)t+\frac{c^2t^3}{3\mathcal{R}^2}-\frac{ct^2}{\mathcal{R}}\theta\right]\right)dt,\nonumber\\
\mathcal{E}_{\perp} & \simeq & \mathcal{E}_o \omega \phi\int_{-\infty}^{\infty}\exp\left(i\frac{\omega}{2}\left[\left(\frac{1}{\gamma^2}+\theta^2+\phi^2\right)t+\frac{c^2t^3}{3\mathcal{R}^2}-\frac{ct^2}{\mathcal{R}}\theta\right]\right)dt.
\end{eqnarray}
We make standard textbook substitutions $x = \frac{ct}{\mathcal{R}}\left(\frac{1}{\gamma^2}+\theta^2+\phi^2\right)^{-1/2}$ and $\xi = \frac{\omega\mathcal{R}}{3c}\left(\frac{1}{\gamma^2}+\theta^2+\phi^2\right)^{3/2}$ to obtain
\begin{eqnarray}
\mathcal{E}_{\parallel} & \simeq & \mathcal{E}_o \omega \frac{\mathcal{R}}{c}\left(\frac{1}{\gamma^2}+\theta^2+\phi^2\right)\int_{-\infty}^{\infty}\left(x-\frac{\theta}{(\frac{1}{\gamma^2}+\theta^2+\phi^2)^{1/2}}\right)\exp\left(i\frac{3}{2}\xi\left(x+\frac{1}{3}x^3-\frac{\theta}{(\frac{1}{\gamma^2}+\theta^2+\phi^2)^{1/2}}x^2\right)\right)dx, \nonumber\\
\mathcal{E}_{\perp} & \simeq & \mathcal{E}_o \omega \frac{\mathcal{R}}{c} \left(\frac{1}{\gamma^2}+\theta^2+\phi^2\right)^{1/2} \phi \int_{-\infty}^{\infty}\exp\left(i\frac{3}{2}\xi\left(x+\frac{1}{3}x^3-\frac{\theta}{(\frac{1}{\gamma^2}+\theta^2+\phi^2)^{1/2}}x^2\right)\right)dx.
\end{eqnarray}
We note that $x+x^3/3-\theta/(\frac{1}{\gamma^2}+\theta^2+\phi^2)^{1/2}x^2\rightarrow x$ for $x\rightarrow0$ and $x+x^3/3-\theta/(\frac{1}{\gamma^2}+\theta^2+\phi^2)^{1/2}x^2\rightarrow x^3/3$ for $x\rightarrow\pm\infty$, and hence we 
can ignore the $x^2$ dependence in the exponential. The remaining integration 
can be expressed in terms of modified Bessel functions $K_{2/3}(\xi)$ and 
$K_{1/3}(\xi)$ giving us the final form of the amplitudes :
\begin{eqnarray}
\mathcal{E}_{\parallel} & \simeq & \gamma \mathcal{E}_o \left(\frac{\omega}{\omega_c}\right) \left[(1 + \gamma^2\theta^2 + \gamma^2\phi^2) K_{2/3}(\xi) - i\gamma\theta(1 + \gamma^2\theta^2 + \gamma^2\phi^2)^{1/2} K_{1/3}(\xi)\right] \nonumber\\
\mathcal{E}_{\perp} & \simeq & i \gamma \mathcal{E}_o \left(\frac{\omega}{\omega_c}\right) \gamma\phi (1 + \gamma^2\theta^2 + \gamma^2\phi^2)^{1/2} K_{1/3}(\xi),
\end{eqnarray}
where $\omega_c = \gamma^3 c/\mathcal{R}$, is the characteristic frequency of 
curvature radiation. The four Stokes parameters can be written in terms of the
two amplitudes as :
\begin{eqnarray}
I & = & \mathcal{E}_{\parallel}^*\mathcal{E}_{\parallel} + \mathcal{E}_{\perp}^*\mathcal{E}_{\perp} \nonumber \\
Q & = &\mathcal{E}_{\parallel}^*\mathcal{E}_{\parallel} - \mathcal{E}_{\perp}^*\mathcal{E}_{\perp} \nonumber \\
U & = & \mathcal{E}_{\parallel}^*\mathcal{E}_{\perp} + \mathcal{E}_{\perp}^*\mathcal{E}_{\parallel} \nonumber \\
V & = & i (\mathcal{E}_{\parallel}^*\mathcal{E}_{\perp} - \mathcal{E}_{\perp}^*\mathcal{E}_{\parallel})
\end{eqnarray}

\end{document}